\documentclass[fleqn,usenatbib]{mnras}

\usepackage{newtxtext,newtxmath}

\usepackage[T1]{fontenc}
\usepackage{ae,aecompl}

\usepackage{graphicx}	
\usepackage{amsmath}	
\usepackage{amssymb}	

\usepackage{enumerate}
\usepackage{multirow}

\usepackage{color}

\newcommand*\mean[1]{\bar{#1}}
\newcommand*\ft[1]{\widetilde{#1}}
\newcommand*\Bell{\ensuremath{{\boldsymbol\ell}}}
\newcommand*\Btheta{\ensuremath{{\boldsymbol\theta}}}
\newcommand*\Bgamma{\ensuremath{{\boldsymbol\gamma}}}
\newcommand*\Bp{\ensuremath{{\bf p}}}
\newcommand*\estp{\ensuremath{\hat{\bf p}}}
\newcommand*\pfid{\ensuremath{{\bf p}_0}}
\newcommand*\pset{\ensuremath{\{{\bf p_i}\}}}
\newcommand*\Bd{\ensuremath{{\bf d}}}
\newcommand*\dobs{\ensuremath{\hat{\Bd}_{\rm obs}}}

\newcommand*\shapecat[2]{\ensuremath{\hat{\Bgamma}#1(\Btheta_g,z_g #2)}}
\newcommand*\scobs{\shapecat{_{\rm obs}}{}}
\newcommand*\screal{\shapecat{_r}{; \Bp}}
\newcommand*\scfid{\shapecat{_r^{fid}}{}}

\newcommand*\shearmap{\ensuremath{\Bgamma(\Btheta_p,\bar{z}_b)}}
\newcommand*\kmap[2]{\ensuremath{\hat{\kappa}#1(\Btheta_p#2)}}

\newcommand*\kobs{\kmap{_{\rm obs}}{}}
\newcommand*\kreal{\kmap{_r}{; \Bp}}
\newcommand*\kfid{\kmap{_r^{\rm fid}}{}}

\newcommand*\ktobs{\kmap{_{\rm obs}}{,\bar{z}_b}}
\newcommand*\ktreal{\kmap{_r}{,\bar{z}_b; \Bp}}
\newcommand*\ktfid{\kmap{_r^{\rm fid}}{,\bar{z}_b}}

\newcommand*\nph[1]{\ensuremath{n_{\rm ph}(#1)}}
\newcommand*\ntrue[1]{\ensuremath{n_{\rm true}(#1)}}
\newcommand*\nest[1]{\ensuremath{\hat{n}(#1)}}

\newcommand*\ntph[1]{\ensuremath{n^i_{\rm ph}(#1)}}
\newcommand*\nttrue[1]{\ensuremath{n^i_{\rm true}(#1)}}
\newcommand*\ntest[1]{\ensuremath{\hat{n}^i(#1)}}

\newcommand*\zph{\ensuremath{z_{\rm ph}}}

\newcommand*\relbias{\ensuremath{\mathcal{B}/\mathcal{U}}}

\newcommand{\gadgettwo}{\mbox{\sc Gadget-2}}
\newcommand{\lenstools}{\mbox{\sc LensTools}}
\newcommand{\camelus}{\mbox{\sc camelus}}
\newcommand{\numpy}{\mbox{\sc numpy}}
\newcommand{\scipy}{\mbox{\sc scipy}}
\newcommand{\pandas}{\mbox{\sc pandas}}
\newcommand{\astropy}{\mbox{\sc astropy}}
\newcommand{\matplotlib}{\mbox{\sc matplotlib}}
\newcommand{\scikitlearn}{\mbox{\sc scikit-learn}}

\defcitealias{petri16b}{P16}
\defcitealias{zorrilla16a}{Z16}
\defcitealias{zorrilla17a}{Z17}

\title[Impact of Photo-$z$ Errors on Lensing Surveys]{The Impact of Photometric Redshift Errors on Lensing Statistics in Ray-Tracing Simulations}

\author[M. W. Abruzzo \& Z. Haiman]{
Matthew W. Abruzzo,$^{1}$\thanks{E-mail: mwa2113@columbia.edu}
and Zolt\'{a}n Haiman$^{1}$
\\
$^{1}$Department of Astronomy, Columbia University, New York, NY 10027, USA
}

\pubyear{2018}

\begin{document}
\label{firstpage}
\pagerange{\pageref{firstpage}--\pageref{lastpage}}
\maketitle

\begin{abstract}
Weak lensing surveys are reaching sensitivities at which uncertainties
in the galaxy redshift distributions $n(z)$ from photo-$z$ errors
degrade cosmological constraints. We use ray-tracing simulations and a simple treatment 
of photo-$z$ errors to assess cosmological parameter biases from uncertainties in $n(z)$ in 
an LSST-like survey. We use the power spectrum and the abundance of 
lensing peaks to infer cosmological parameters, and find that the 
former is somewhat more resilient to photo-$z$ errors. We place conservative lower 
limits on the survey size at which different types of photo-$z$ errors 
degrade $\Lambda$CDM ($w$CDM) parameter constraints~by~$50\%$. A
residual constant photo-$z$ bias of $|\delta z|<0.003(1+z)$, 
satisfying the current LSST requirement, does not significantly degrade 
constraints for surveys smaller than $\approx$$1300$ ($\approx$$490$) 
deg$^2$ using lensing peaks and $\approx$$6500$ ($\approx$$4900$) deg$^2$ 
using the power spectrum. Adopting a recent prediction for LSST's full 
photo-$z$ probability distribution function (PDF), we 
find that simply approximating $n(z)$ with the photo-$z$ galaxy distribution 
directly computed from this PDF would degrade surveys as 
small as $\approx$$60$ ($\approx$$65$) deg$^2$ using lensing peaks or 
the power spectrum. Assuming that the centroid bias in each tomographic redshift 
bin can be removed from the photo-$z$ galaxy distribution, using lensing peaks 
or the power spectrum still degrades surveys larger than $\approx$200 
($\approx$255) or $\approx$248 ($\approx$315) deg$^2$. These results imply that 
the expected broad photo-$z$ PDF significantly biases parameters, which needs to 
be further mitigated using more sophisticated photo-$z$ treatments.
\vspace{2.5\baselineskip}
\end{abstract}




\section{Introduction}

Upcoming weak lensing (WL) surveys hold great promise for probing the 
nature of dark energy and constraining cosmological parameters to 
unprecedented precision. They draw their constraining power from 
measuring cosmic shear -- small distortions in the shapes of distant 
galaxies caused by gravitational lensing by foreground structures -- 
which is sensitive to the histories of both the expansion rate and 
the growth of structure.

Current WL surveys are reaching sizes where the constraints on
$\Omega_m$ and $\sigma_8$ are becoming competitive with other
cosmology probes. Measurements of the shear 2-point correlation
function (2PCF) for the recently completed
Canada-France-Hawaii-Telescope Legacy
Survey\footnote{\url{http://www.cfhtlens.org}} (CFHTLenS)
used 4.2 million galaxies distributed over 154 ${\rm deg}^2$
\citep{kilbinger13a}. Constraints have also been
measured in ongoing surveys such as the Kilo Degree
Survey\footnote{\url{http://kids.strw.leidenuniv.nl}} (KiDS)
\citep{dejong13a}, the Dark Energy
Survey\footnote{\url{http://www.darkenergysurvey.org}}
(DES) \citep{des05a}, and Subaru Hyper 
Suprime-Cam\footnote{\url{https://hsc.mtk.nao.ac.jp/}} (HSC)
\citep{aihara18a}. Current measurements of two-point statistics 
in each survey have used 15 million galaxies distributed over 
450 ${\rm deg}^2$ \citep{hildebrandt17a}, 26 million galaxies
distributed across 1321 ${\rm deg}^2$ \citep{troxel17a}, and 9 
million galaxies distributed over 137 ${\rm deg}^2$ \citep{hikage18a}. 
Upon completion these surveys will span 1500 ${\rm deg}^2$, 
5000 ${\rm deg}^2$, and 1400 ${\rm deg}^2$, respectively.

Moreover, when combined with other cosmological probes, such as the
cosmic microwave background or galaxy clustering, current WL surveys
are beginning to provide useful constraints on $w$
\citep[e.g.][]{des17a}.  Data from current surveys has also been used
compute non-Gaussian summary statistics, such as lensing peaks
\citep[e.g][]{liu15a,liu15b,kacprzak16a,martinet18a}, higher-order
mixed moments \citep[e.g.][]{petri15a}, higher-order correlation
functions \citep[e.g.][]{fu14a}, and Minkowski functionals
\citep[e.g][]{petri15a}. These summary statistics provide the
opportunity to increase the amount of cosmological information
extracted from cosmic shear. More recently, studies have explored
possible additional features residing in simulated lensing maps, using
convolutional neural networks \citep[e.g.][]{schmelzle17a, gupta18a,
  ribli18a}.

Upcoming WL surveys such as the full DES,
Euclid\footnote{\url{http://sci.esa.int/euclid}} \citep{laureijs11a},
LSST\footnote{\url{http://www.lsst.org}} \citep{ivezic08a, lsst09a},
and WFIRST\footnote{\url{https://wfirst.gsfc.nasa.gov}}
\citep{spergel15a}, are expected to measure cosmic shear using $\sim
10^8-10^9$ galaxies. Cosmological constraints from these
surveys are likely to be limited by systematic errors \citep{albrecht06a}. 
One example is the systematic error arising from uncertainties and biases 
in the photometric redshift (photo-$z$) measurements.

Due to the prohibitive cost of acquiring spectroscopic redshifts for
every galaxy, weak lensing surveys rely on photo-$z$'s to estimate the
underlying true galaxy redshift distribution $n_{\rm true}(z)$.
Because photo-$z$'s are estimated with a finite set of photometric
filters, their errors vary as functions of $z$, brightness, and
morphological type. Typically these errors are divided into bias (the
offset between the median/mean photo-$z$ and the true $z$), the
scatter of photo-$z$ measurements about the median photo-$z$ at a
given $z$, and the catastrophic photo-$z$ errors (outlier photo-$z$'s
that can arise from multimodal photo-$z$ distributions). 
On the observational side, WL surveys perform complex analyses to reduce 
errors in estimates of \ntrue{z}. However, theoretical work is required 
to understand how small errors in these estimates propagate into 
cosmology parameter errors.

Extensive work has been done to study the impact of photo-$z$ errors
on constraints from WL surveys and techniques to mitigate it
\citep[e.g.][]{ma06, huterer06a, brindle07a, jain07a, abdalla08a,
  ma08, kitching08a, bernstein10a, hearin10a, hearin12a, cunha12a,
  cunha14a, dePutter14a, shirasaki14a, petri16b, rau17a}. The vast
majority of works used a Fisher analysis to study the impact of
photo-$z$ errors on either the lensing 2PCF or the power spectrum, and
on the cosmological parameters inferred from these
observables. Notable exceptions include \citet{huterer06a} who studied
the impact on the convergence bispectrum, \citet{kitching08a} who
exclusively studied the impact on constraints obtained with the
shear-ratio method and 3D cosmic shear, \citet{shirasaki14a} who
employed ray-tracing simulations to assess the impact on Minkowski
Functionals and \citet{petri16b} who used ray-tracing simulations to
study the impact on both peak counts and higher-order mixed moments of
the convergence field. Because they have
different dependence on the underlying galaxy distributions,
non-Gaussian summary statistics are impacted in different ways by
photo-$z$ errors and offer the possibility of self-calibration
\citep{huterer06a,petri16b}.

Ray-tracing simulations have also been used to assess the effects of
other kinds of systematics on cosmological constraints of
non-Gaussian Statistics
\citep[e.g.][]{shirasaki13a,liu14a,petri14a}.

In this paper, we use ray-tracing simulations, without assuming a
linear dependence of summary statistics on cosmology, to study the
impact of both a constant uncalibrated photo-$z$ bias and of a more
realistic full photo-$z$ probability distribution function (PDF;
adapted from \citealt{rhodes17a}, based on a simulated
spectroscopic calibration sample for LSST).
Our approach is to use our simulations to produce a mock
lensing dataset; we then produce an independent set of simulations,
over a large grid of cosmologies, which all have the ``wrong'' redshifts,
and which we use to fit the mock data. We quantify the resulting
degradation in the cosmological parameter constrains inferred from the
tomographic convergence peak counts in an LSST-like survey. We also
present a side-by-side analysis of the tomographic convergence power
spectrum, to compare the susceptibility of these two observables to
photo-$z$ errors.

We organise this paper as follows: In \S~\ref{section:methods} we
describe our simulations, the construction of the convergence maps,
the calculation of the summary statistics, our model of the photo-$z$
errors, and scaling up our forecast degradations to large LSST-like
surveys.  We then describe and discuss our results in
\S~\ref{section:results} and \S~\ref{section:discussion},
respectively. Finally, in \S~\ref{section:conclusion} we summarise our
main conclusions and the implications of this work.

Our analysis is the first step towards a full assessment, based on
simulations, to identify photo-$z$ requirements for peak counts, and
to design the best strategy to mitigate the impact of photo-$z$
errors.   

\begin{table*}
  \centering
  \caption{ This table summarises details
    of the Single-$z$ dataset and Tomographic dataset. We define \kobs,
    \kfid\ and \kreal\ as the as the convergence maps used as mock
    observations, to construct the covariance matrix, and as
    theoretical predictions for inference, respectively.  The first
    two are evaluated in the fiducial cosmology, while the last are
    evaluated for different combinations of cosmological parameters
    $\Bp$.  Corresponding shear catalogues are denoted by \scobs,
    \scfid\ and \screal.  Note that $\Btheta_p$ indicates the location
    of a pixel on a two-dimensional convergence map, while $\Btheta_g$
    and $z_g$ indicate the apparent angular position and redshift of a
    galaxy in a shape catalogue. } 
  \label{tab:sim_overview}
  \begin{tabular}{p{2.1in}|p{2.1in}p{2.1in}}
    \hline
     & Single-$z$ dataset & Tomographic dataset \\
    \hline
    \gadgettwo\ comoving volume &
    $(240\ h^{-1}{\rm Mpc})^3$ &
    $(260\ h^{-1}{\rm Mpc})^3$
    \\
    Comoving lens plane spacing &
    $80\ h^{-1}{\rm Mpc}$ &
    $120\ {\rm Mpc}$
    \\
    Simulations$^a$: \kobs\ \& \scobs &
    \citetalias{zorrilla17a} &
    Simulations were run for this work
    \\
    Simulations$^a$: \kfid\ \& \scfid &
    Same simulation used for \kobs &
    \citetalias{petri16b}
    \\
    Simulations$^a$: \kreal\ \& \screal &
    161 of the sampled cosmologies from \citetalias{zorrilla16a} \& simulation
    for \kobs &
    All 100 cosmologies from \citetalias{petri16b} [does not include simulation
    for \scfid]
    \\
    photo-$z$ galaxy distribution & $\nph{z}=25\,{\rm arcmin}^{-2}\delta(z-1)$ &
    Equation~\ref{eqn:gal_dist}
    \\
    Number of realisations ($N_r$) &
    500 &
    16000 [for \screal\ \& \scfid] and 1000 [for \scobs]
    \\
    Tomography & No &
    Yes - see Fig.~\ref{figure:reshift_bin} and Table~\ref{tab:tomo_details} 
    \\
    $\kappa$ map Gaussian smoothing scale &
    $\sqrt{0.5}$ arcmin &
    1 arcmin
    \\
    $\kappa$ map resolution &
    $1024^2$ & $492^2$
    \\
    $\kappa$ map Field of View &
    $(3.5{\rm deg})^2$ &
    $(3.36{\rm deg})^2$
    \\
    Summary statistics &
    $n_{\rm pk}$ &
    $n_{\rm pk}$ and $P^{\kappa\kappa}$
    \\
    PCA & No & Only for $P^{\kappa \kappa}$
    \\
    Role of interpolator &
    Interpolate the $\chi^2$ between the observation and each
    sampled cosmology &
    Interpolate the summary statistics between each sampled
    cosmology
    \\
    Inferred Cosmological Models &
    $\Lambda$CDM &
    $\Lambda$CDM and $w$CDM
    \\
    Uniform Prior Range &
    $\Omega_m \in [0.16,0.6]$, $\sigma_8\in[0.15,1.25]$ &
    $\Omega_m \in [0.2,0.5]$, $w \in [-1.5,-0.5]$ (for $w$CDM inference),
    $\sigma_8\in[0.5,1.2]$
    \\
    Sampling of Inference Grid &
    $\Delta p = 0.001$ in each parameter \citepalias{zorrilla16a} &
    See Table~\ref{tab:cosmo_models}
    \\
    \hline
  \end{tabular}
  \\$^a$ Indicate the works from which the simulations 
  	used to construct different convergence and shear maps are 
    adopted.
\end{table*}

\section{Methodology}\label{section:methods}

In our analysis, we use ray-tracing simulations coupled with N-body 
DM-only simulations to construct pixelized convergence ($\kappa$) maps. 
The convergence is proportional to the distance-weighted over-density 
along the light's path. The $\kappa$ maps are then used to compute 
weak lensing summary statistics \Bd. Our basic approach is to produce 
these statistics in a fiducial cosmology \pfid, representing a mock 
observation. \pfid\ is a $\Lambda$CDM cosmology 
consistent with WMAP results \citep{hinshaw13a}, with the parameter 
values $(h,\Omega_m,\Omega_\Lambda,\Omega_b,w,\sigma_8,n_s)=(0.72,$ $0.26,$
$0.74,$ $0.0046,$ $-1,$ $0.8,$ $0.96)$. We then separately
produce a different set of the maps and statistics, over a large grid of
spatially flat cosmological models, \pset; these represent the theoretical 
predictions used to fit the mock observation and to infer
cosmological parameters $\Bp \equiv (\Omega_m,w,\sigma_8)$. The mock 
observation and the theoretical predictions have different 
redshift-distributions, which results in biases in the best-fit 
parameters.

Our analysis involves the construction of $N_r$ realisations of three
types of pixelized convergence maps: \ktobs, \ktfid\ and \ktreal. To be 
consistent with the notation of \citet{petri16b}, $\hat{\kappa}$ denotes 
convergence maps with shape noise added while $\Btheta_p$ and $\mean{z}_b$ 
indicate the coordinates of a pixel and the tomographic redshift bin, 
respectively. \ktobs\ refers to the maps constructed from \pfid\ which we use 
to produce summary statistics for our the mock observation. \ktfid\ is also 
constructed for \pfid, however we use the calculated \Bd\ to compute a covariance 
matrix over its $N_r$ realisations. This is necessary to evaluate parameter 
likelihoods. Finally, we construct \ktreal\ for each cosmology in \pset\ and 
use it compute the expectation value of the summary statistics $\Bd(\Bp)$ 
(again over $N_r$ realisations). We use them as a theoretical prediction tool that
interpolates either the \Bd\ themselves or the goodness of fit of a mock observation 
as a function of cosmology. 

To assess the impact of photo-$z$ errors, we use a single galaxy 
distribution $n(z)$ to produce the covariance matrix and interpolator, and 
examine the biases in the constraints for mock observations computed with 
different galaxy distributions. These distributions represent different 
types of photo-$z$ errors, including a 
residual photo-$z$ bias, and a more realistic full photo-$z$ PDF simulated 
for LSST. We quantify the degradation in the cosmological parameter 
constrains inferred from the tomographic $\kappa$ peak counts in an 
LSST-like survey. We also present a side-by-side analysis of the tomographic 
$\kappa$ power spectrum, to compare the susceptibility of these two 
observables to photo-$z$ errors.

In \S~\ref{section:cs_sim} and
\S~\ref{section:kappa_prep}, we review the steps to simulate cosmic
shear for an arbitrary galaxy distribution in an arbitrary cosmology
$\Bp$, and to prepare convergence ($\kappa$) maps suitable for computing 
summary statistics.  We describe the steps to calculate summary statistics in
\S~\ref{section:summary_statistics}. In \S~\ref{section:photoz_model}, 
we describe how we model photo-$z$ errors, and finally, in 
\S~\ref{section:inference} we describe cosmological parameter inference.

Our analysis makes use of two separate datasets, summarised in
Table~\ref{tab:sim_overview}, mostly consisting of simulation data we
adopt from prior studies. Each dataset has an independent \pset.  Our
first dataset includes data from \citet[][hereafter Z16]{zorrilla16a}
and \citet[][Z17]{zorrilla17a}. \citetalias{zorrilla16a} studied the accuracy
of using lensing peak predictions made with \camelus\ \citep{lin15a}
to infer cosmology, while \citetalias{zorrilla17a} studied the individual
and combined dependence of weak lensing on the growth of structure and
on the universe's expansion history. Both of these studies used source
galaxies at a single redshift ($z=1$), and did not include redshift
tomography. We will hereafter refer to the combined dataset as the
``Single-$z$'' dataset. We also remove the redshift-bin dependence
from the notation in this dataset, and denote \ktreal, \ktfid, and
\ktobs\ as \kreal, \kfid, and \kobs.

For the Single-$z$ dataset, we directly take the $\Bd$ computed for
161 of our 162 cosmologies in \pset\ for \ktreal\ from
\citetalias{zorrilla16a}.  The remaining cosmology in \pset\ is
\pfid. We reuse the single set of lens planes for \pfid\ from 
\citetalias{zorrilla17a} to construct \kmap{_r}{; \pfid}, \kfid, and 
\kobs. This is because the lens planes in the fiducial model of 
\citetalias{zorrilla16a} were no longer available, and we need 
access to these lens planes in order to perform new ray-tracing 
calculations in the same model to slightly different redshifts, in 
the presence of photo-$z$ errors (see below).

The second, hereafter the ``Tomography'', dataset almost entirely 
consists of data from \citet[][P16]{petri16b}. \citetalias{petri16b} 
forecasted cosmology constraints for a survey with an LSST-like galaxy 
distribution using the tomographic convergence power spectrum, tomographic peak counts
and nine mixed moments for each of the tomographic convergence maps. 
Additionally, they made use of Principal Component analysis and briefly 
assessed the impact of uncorrected photo-$z$ errors on cosmological 
constraints.  Our use of redshift tomography in this case necessitates 
the intermediate step of producing shape catalogues \screal, \scfid, and 
\scobs\ from which we ultimately construct \ktreal, \ktfid, and \ktobs. We 
call a noiseless catalogue a ``Shear Catalogue'' but adopt the convention of 
``Shape Catalog'' in the presence of shape noise. For this dataset, we reuse the
shear catalogues from \citetalias{petri16b} to generate \ktreal\ and \ktfid. To 
generate \ktobs, we use new simulation data specifically generated for this work.

For completeness and convenience, we summarise the numerous small 
differences between the datasets in Table~\ref{tab:sim_overview}.

\begin{figure}
  \center
\includegraphics[width = 3.5 in]{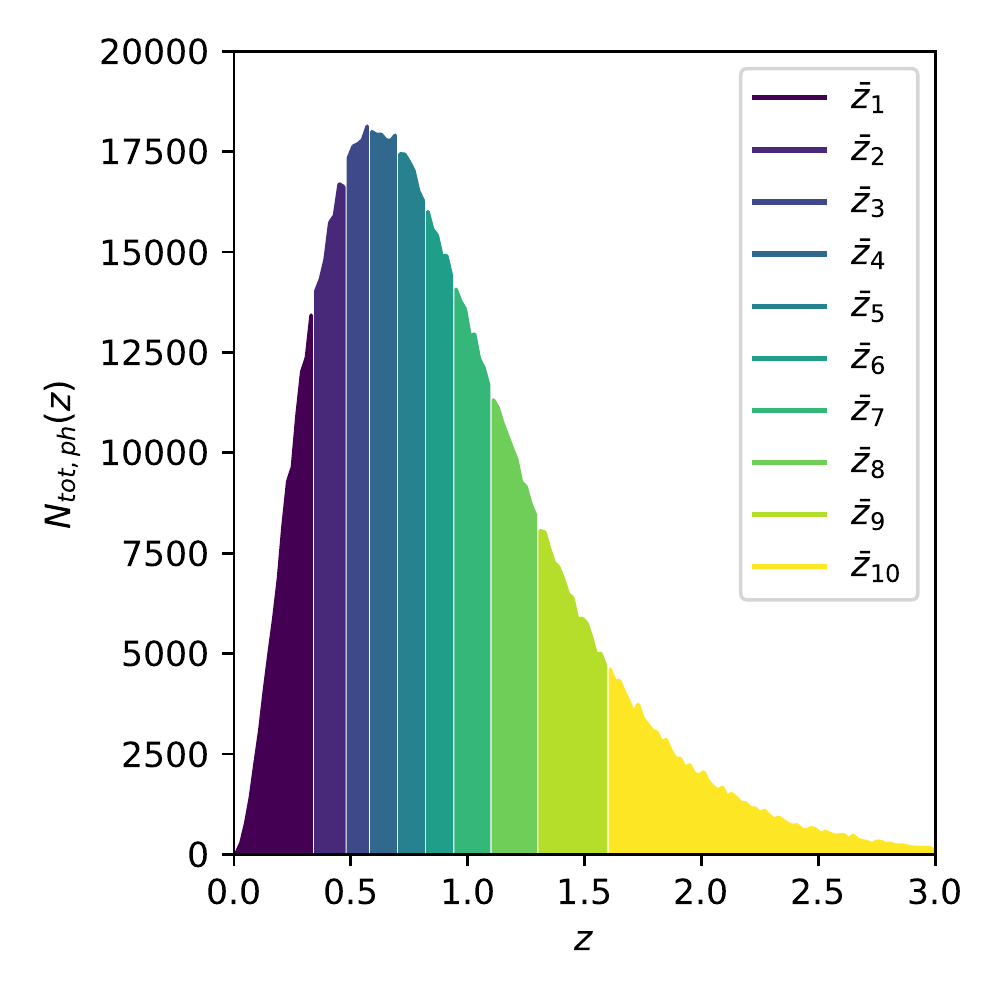}
\caption{\label{figure:reshift_bin} Illustrates the photo-$z$
  distribution and the tomographic bins. The 10 bins are chosen
  to cover the range $0\leq\zph\leq3$, and to each contain $10^5$
  galaxies per simulated (3.5 deg)$^2$ field.}
\end{figure}

\subsection{Cosmic shear simulations} \label{section:cs_sim}
In this subsection, we review how to simulate cosmic shear.

For a specific cosmology, \Bp, we start by
running a \gadgettwo\ \citep{springel05a} dark matter-only N-body
simulation. For the Single-$z$ (Tomographic) Dataset we use a box
of comoving side length $L_b=240\ (260)\ h^{-1}{\rm Mpc} $
that contains $512^3$ DM particles with mass resolution of
$\approx 10^{10} {\rm M}_{\sun}$ per particle. We slice each snapshot
and apply random shifts and rotations to construct two-dimensional 
lens planes with comoving thickness $80\ h^{-1}{\rm Mpc}$ ($120$ Mpc).
Next, we line up the planes perpendicular to the observer's 
line of sight and calculate the induced $\kappa$ or shear $\Bgamma$ for a source at redshift
$z_s$, arising from deflections to the light rays by the sequence of
lenses. We employ a multi-plane ray-tracing
algorithm \citep{jain00a,hilbert09a} to trace the path of light rays
between the angular position at $z_s$ and its apparent angular position
on~the~sky~$\Btheta$ at $z=0$. In practice, we use the
\lenstools\ package \citep{petri16c} to carry out all of these
calculations for collections of sources distributed in angular
position $\Btheta$ and redshift $z_s$.

Throughout our analysis, we hold the distribution of the apparent
galaxy sky positions constant and redo ray-tracing with varied galaxy
redshift distributions.  For the Single-$z$ (Tomographic) dataset, we
assume that the galaxies are uniformly distributed across a 
$(3.5{\rm deg})^2$ region with an average number density $n_g$=25 
${\rm arcmin}^{-2}$ (22 ${\rm arcmin}^{-2}$). Due to computational
limitations, we use a constant photo-$z$ galaxy distribution \nph{z}
for each simulation dataset and use it to act as the distribution of
source redshifts for the majority of our ray-tracing simulations. This
choice allows for the reuse of the summary statistic measurements from
\citetalias{zorrilla16a} and the shear catalogues from 
\citetalias{petri16b}. We defer further discussion of the
interpretation of the fixed \nph{z} to \S~\ref{section:photoz_model}.

In the Single-$z$ dataset, we assume a galaxy surface density redshift
distribution,  $d^2N_{tot}/(d\Omega dz_{\rm ph})$, referred to as \nph{z} 
for simplicity, given by $\nph{z}=25 {\rm arcmin}^{-2} \delta(z-1)$.
In contrast, the analysis of the Tomographic dataset assumes a discrete 
distribution of $N_{tot}=10^6$ galaxies with photo-$z$'s restricted to 
\zph$\leq 3$ and otherwise drawn from
\begin{equation}
  \label{eqn:gal_dist}
  \nph{z}=n_0\left(\frac{z}{z_0}\right)^2 {\rm
    exp}\left(-\frac{z}{z_0}\right).
\end{equation}

In this equation, $z_0=0.3$ was chosen to match the expected
spectroscopic distribution for LSST and $n_0$ is chosen such that
$N_{tot,ph}(z)=(3.5{\rm deg})^2\times \nph{z}$ integrates to $N_{tot}$.
Figure~\ref{figure:reshift_bin} illustrates $N_{tot,ph}(z)$ for the
Tomographic dataset and the 10 tomographic bins we employ in our
analysis; each bin contains $10^5$ galaxies. Each of these distributions 
are identical to the spectroscopic redshift distributions \ntrue{z} used 
for ray-tracing in \citetalias{zorrilla16a} and \citetalias{petri16b}.

Because the Single-$z$ dataset employs a simple distribution of
galaxies, we are able to directly produce a $\kappa$ map covering a
$(3.5{\rm deg})^2$ field with a resolution of $1024^2$ pixels from ray-tracing.
To model the impact of shape noise, we
follow \citet{vanWaerbeke00a} and \citetalias{zorrilla16a}, and add a 2D array
of values drawn from a zero-mean Gaussian distribution with standard deviation
\begin{equation}
  \sigma = \sqrt{\frac{\sigma_\epsilon^2}{2n_g A_{pix}}}.
\end{equation}
In this equation, $\sigma_\epsilon = 0.4$
corresponds to the r.m.s intrinsic ellipticity,
$n_g=25\,{\rm arcmin}^{-2}$
is the galaxy surface density and $A_{pix}=(0.042\,{\rm arcmin})^{2}$
is the solid angle enclosed by a pixel in the convergence map.

Ray-tracing for the Tomographic dataset produces a shear catalogue $\{
\Bgamma_g \}$ which includes the induced shear on each galaxy. We
model the impact of shape noise as in \citetalias{petri16b}. In a real
survey we measure galaxy ellipticity $\Bgamma_m = \Bgamma +
{\boldsymbol \epsilon}_{\rm intrinsic}$, where $\Bgamma$ is the true
cosmic shear and ${\boldsymbol \epsilon}_{\rm intrinsic}$ is the
intrinsic ellipticity of the galaxy
\citep{schneider05a}. Consequently, we transform our shear catalogue
into a shape catalogue, which lists the measured ellipticity of each
galaxy, by modelling ${\boldsymbol \epsilon}_{\rm intrinsic}$ with
randomly drawn values from a zero-mean Gaussian with standard
deviation $\sigma(z) = 0.15 + 0.035 z_s$ \citep{song04a}.

By randomising the slices, shifts, and rotations used to
create the lens planes and changing the random seed used for 
modelling shape noise, we can generate pseudo-random
realisations of convergence maps and shape catalogues. 

For the Single-$z$ dataset, we reuse lensing peaks for all cosmologies in 
\pset\ other than \pfid\ from \citetalias{zorrilla16a}. These lensing peaks were 
computed from \kreal\ which had been constructed using the above procedure. Because we 
want our mock observations, in the absence of photo-$z$ errors, to have $\chi^2=0$ 
when compared to the model-predicted lensing peaks, the lensing peaks computed from 
$\kmap{_r}{; \pfid}$, we need to use the same initial conditions and random seed to 
produce realisation $r$ of $\kmap{_r}{; \pfid}$ and \kobs. Additionally, to produce 
the \kobs\ for different sets of photo-$z$ errors, we need access to the lens planes 
and random seed information. Since neither are available from \citetalias{zorrilla16a}, 
we use the lens planes for \pfid\ from \citetalias{zorrilla17a} and a different seed from 
\citetalias{zorrilla16a} to produce realisation $r$ $\kmap{_r}{; \pfid}$, \kfid, and 
\kobs. Our analysis of this dataset uses the summary statistics 
computed from $N_r=500$ realisations of each type of convergence map.

Our analysis using the Tomographic Dataset employs the suite
of $N_r=16000$ realisations of \screal\ and \scfid\ directly
produced, without shape noise, by \citetalias{petri16b}. For this
dataset, \pset\ includes $P=100$ different cosmologies and does not
include \pfid. The generation of \scfid\ made use of 5 independent
N-body simulations with initial conditions independent from the
$P=100$ other simulations. For more details, we refer the reader to
\citetalias{petri16b}. To produce \scobs, we perform ray-tracing using
lensing planes mixed from 2 independent N-body simulations newly run for
the present work.\footnote{These lens planes are constructed at 
$z$ halfway between the $z$ of the planes used in the 
creation of \screal\ and \scfid; we believe this has a 
negligible impact on our results.} We only ever produce 
$N_r=1000$ realisations of \scobs, and note that realisation $r$ of 
\screal, \scfid, and \scobs\ all use the same random seed for modelling 
shape noise.

\subsection{Convergence Map Preparation} \label{section:kappa_prep}

We next discuss the steps required to transform the
ray-tracing products into convergence maps appropriate for the
calculation of summary statistics.

For the Tomographic dataset, we follow the procedure described by
\citetalias{petri16b}, to construct square shear maps, for each tomographic
bin, with 512 pixels per side and covering an angular area of
$(3.5{\rm deg})^2$ from a shear catalogue $\Bgamma(\Btheta_g,z_g)$. 
Recall that $\Btheta_g$ and $z_g$ indicate the location and redshift of 
galaxy $g$. The
value of the resulting shear map, \shearmap, at pixel $\Btheta_p$ of
tomographic bin $\bar{z}_b$ is given by
\begin{eqnarray}
  \shearmap = \frac{\sum_{g=1}^{N_g} \Bgamma(\Btheta_g,z_g)
    W(\Btheta_g, \Btheta_p; z_g, \bar{z}_b)}
            {\sum_{g=1}^{N_g} W(\Btheta_g, \Btheta_p; z_g, \bar{z}_b)} \\
  W(\Btheta_g, \Btheta_p; z_g, \bar{z}_b) =
    \begin{cases}
      1 &\text{if $\Btheta_g\in\Btheta_p,z_g\in\bar{z}_b$}\\
      0 &\text{otherwise}
    \end{cases}.
\end{eqnarray}
Like \citetalias{petri16b}, we set all pixels of the \shearmap\ without
any galaxies to 0; there are $\sim0.38$ galaxies for every pixel.

Following \citet{kaiser93a}, the Fourier transform of the convergence
map, $\ft{\kappa}(\Bell, \mean{z}_b)$, is given by the E-mode of
the shear map
\begin{equation}
  \label{eqn:ks}
  \ft{\kappa}(\Bell, \mean{z}_b) =
  \frac{(\ell_x^2-\ell_y^2)\ft{\gamma}_1(\Bell, \mean{z}_b) + 2
    \ell_x\ell_y \ft{\gamma}_2(\Bell, \mean{z}_b)}{\ell_x^2+\ell_y^2}.
\end{equation}
The inverse Fourier transform of Equation~\ref{eqn:ks} yields the
convergence map $\kappa(\Btheta_p,\mean{z}_b)$. For both datasets, 
the final step is to smooth the convergence with a Gaussian Filter. For the Single-$z$ and
Tomographic datasets, the filters have standard deviations of $\sqrt{0.5}$ and $1$ arcmin. The former is
the same scale used by \citetalias{zorrilla16a}\footnote{Our reuse of 
summary statistics computed by \citetalias{zorrilla16a} requires us to 
replicate their smoothing scale.} while the 
latter is twice the scale used by \citetalias{petri16b}.

The Kaiser-Squires transform and smoothing are both convolution
operations that require assumptions about boundary conditions. We
refer the reader to Appendix~\ref{app:bc} for a brief analysis of this
issue. We conclude that the boundary condition is insignificant for
the Kaiser-Squire transform and clip the 10 outermost pixels of all
$\kappa$ maps, produced for the Tomographic dataset, to completely
eliminate the edge effects from smoothing. However, for the Single-$z$
dataset, we do not discard any pixels.

We apply the relevant preparatory steps, as discussed above, during
the generation of \kreal, \kfid\ and \kobs\ of the Single-$z$ dataset,
and to convert \screal, \scfid\ and \scobs\ of the Tomographic dataset
into \ktreal, \ktfid, and \ktobs. The final maps of the
Single-$z$ (Tomographic) dataset have 1024 (492) pixels per side and
enclose 12.25 (11.29) deg$^2$.

\subsection{Summary Statistics}\label{section:summary_statistics}
\begin{table}
  \centering
  \caption{ Tomographic redshift bin details and the $\nu$ ranges
    used for the peak counts. These ranges are chosen to allow for up to 30 equally sized 
    $\nu$ bins while ensuring that the covariance matrix has non-zero diagonal terms.}
  \label{tab:tomo_details}
  \begin{tabular}{ccc}
    \hline
    bin & $z$ Range & $\nu$ Range \\
    \hline
    $\bar{z}_1$    & $[0,0.332)$      &  $[-3, 6.5)$ \\
    $\bar{z}_2$    & $[0.332, 0.464)$ &  $[-3, 8.5)$ \\
    $\bar{z}_3$    & $[0.464, 0.577)$ &  $[-3, 10.5)$ \\
    $\bar{z}_4$    & $[0.577, 0.689)$ &  $[-3, 10.25)$ \\
    $\bar{z}_5$    & $[0.689, 0.806)$ &  $[-3, 10.5)$ \\
    $\bar{z}_6$    & $[0.806, 0.936)$ &  $[-3, 11)$ \\
    $\bar{z}_7$    & $[0.936, 1.089)$ &  $[-3, 10.5)$ \\
    $\bar{z}_8$    & $[1.089, 1.287)$ &  $[-3, 11.25)$ \\
    $\bar{z}_9$    & $[1.287, 1.596)$ &  $[-3, 10.25)$ \\
    $\bar{z}_{10}$  & $[1.596, 3]$    &  $[-3, 9.5)$ \\
    \hline
  \end{tabular}
\end{table}

From the $\kappa$ maps of the Single-$z$ dataset, we compute the peak
counts $n_{\rm pk}(\nu)$. For the Tomographic dataset, we compute two
summary statistics: (1) tomographic peak counts $n_{\rm
  pk}(\nu,\mean{z}_b)$ and (2) the tomographic power spectrum
$P^{\kappa\kappa}(\ell,\mean{z}_b,\mean{z}_{b^\prime})$. We follow the notation in  \citetalias{petri16b} and define $\Bd(\Bp)$ as the expectation value of a summary statistic in
cosmology \Bp. We can calculate it by averaging the summary statistics
computed from \kreal\ or \ktreal\ over all realisations $r=1, ...,
N_r$. Additionally, we define the observed summary statistic \dobs\ as the average of this statistic in all realisations of
\kobs\ or \ktobs.

Tomographic peak counts $n_{\rm pk}(\nu,\mean{z}_b)$ are
defined as the histogram of $\nu$, the signal-to-noise, of all of the
local maxima in a $\kappa$ map constructed from observed galaxies in
tomographic bin $\mean{z}_b$.  We define the $\nu$ of a given peak in
a $\kappa$ map of bin $\mean{z}_b$ as $\kappa_{\rm
  peak}/\mean{\sigma}_b$ where
$\kappa_{\rm  peak}$ is the convergence in the peak pixel, and
$\mean{\sigma}_b$ is the standard deviation (measured 
over all realisations or in individual maps; see below).
Note that $\mean{\sigma}_b$ is constant for the calculation of all peak counts in a given bin $\mean{z}_b$ of a given dataset.  The 
definition of $n_{\rm pk}(\nu)$ is identical to $n_{\rm pk}(\nu,\mean{z}_b)$ using a single tomographic bin.

The preceding definition describes ``unscaled'' peak counts. Scaled 
tomographic peak counts, which are used by \citetalias{petri16b}, are 
the same in all respects, except that instead of defining $\nu$ with a 
constant $\mean{\sigma}_b$, $\kappa_{\rm peak}$ is divided by the standard 
deviation of the $\kappa$ map for which you are computing the peak counts 
\citep{yang11}. Doing this ``scales out'' the cosmological information 
carried in the standard deviation of the convergence map, which is already 
measured by the power spectrum.

As in \citetalias{zorrilla16a}, we use 100 equally-sized $\nu$ bins
distributed over $-2.0\leq \nu \leq 6.0$ for the Single-$z$ dataset.
For the Tomographic dataset, we use 10 equally spaced $\nu$ per 
$\mean{z}_b$, spanning ranges of $\nu$ listed in Table~\ref{tab:tomo_details}.
We find negligible improvements in our constraints if we use 30 $\nu$ bins per 
$\mean{z}_b$, over the same ranges.

We adopt the same definition for the tomographic power spectrum
$P^{\kappa\kappa}(\ell,\mean{z}_b,\mean{z}_{b^\prime})$
as \citetalias{petri16b}:
\begin{equation}
  \label{eqn:ps}
  \left<\ft{\kappa}(\Bell, \mean{z}_b) \ft{\kappa}(\Bell^\prime,
  \mean{z}_{b^\prime})\right> \equiv
  (2\pi)^2\delta_D(\Bell-\Bell^\prime)
  P^{\kappa\kappa}(\ell,\mean{z}_b,\mean{z}_{b^\prime}),
\end{equation}
where the angular brackets indicate the average over all orientations of 
the wavenumber of length $\ell=\ell^\prime$.
Similar to \citetalias{petri16b}, we use 15 uniformly sized multipole
bands spanning $200<\ell<2000$, compute all auto-correlation spectra,
and compute cross-spectra between all unique combinations of
tomographic bins.

We measure $100$ components of the $n_{\rm pk}(\nu)$ For the Single-$z$ 
dataset.  For the Tomographic dataset, the $n_{\rm pk}(\nu,\mean{z}_b)$ has 
$10\,(\rm{tomographic\ bins})\times 10\,(\nu\ {\rm bins})=100$
components while the $P^{\kappa\kappa}(\ell,\mean{z}_b,\mean{z}_{b^\prime})$
consists of $45\,({\rm unique\ cross})+10\,({\rm auto\ correlated}) = 55$ 
spectra and a total of $15\,({\rm multipoles})\times 55\,({\rm spectra})=825$ 
multipole bands.

\subsection{Modelling photo-$z$ errors} \label{section:photoz_model}

In this section, we describe how we model the impact of photo-$z$
errors.  In short, our approach is to simulate a mock observation
[\kobs\ or \ktobs], in which we ray-trace to redshifts $z^\prime\neq
z$, slightly offset from the original redshifts $z$ used in our suites
on which the predictions $\Bd(\Bp)$ are based [\kreal\ and \ktreal].
This mock observation represents the true universe.  We then fit this
observation with $\Bd(\Bp)$ created with the
original redshifts $z$.  In this approach, $z^\prime$ plays the role
of the true redshift, and $z$ plays the role of the redshifts assigned
to galaxies in the observation, based on photo-$z$'s (either directly
the photometric redshift, or a calibrated/corrected version). In
general, this yields a best-fit cosmology \estp\ that is biased
and also modifies the shape of the inferred
confidence contours (see below).

This approach -- switching the role of the true and the
observationally estimated redshifts -- has a shortcoming: it allows
only one fixed set of redshifts to be assigned to galaxies,
i.e. fixing $\hat{n}(z)$. The major advantage is that this requires
only one ray-tracing calculation for a given photo-$z$ distribution.
In interpreting a real observation, one would simultaneously fit for
the unknown cosmological parameters and the unknown true
redshift distribution $n_{\rm true}(z)$.  This would require repeated
ray-tracing, and re-computing the predictions for the observables
\Bd\ for each hypothesised $n(z)$ in each test cosmology.  This is
beyond the capability of our current emulator, and will need to be
addressed in future work.

Typically, photo-$z$ calculation techniques are trained using
large spectroscopic calibration samples, and when applied to
galaxies in a survey they yield either a point estimate of redshift,
$z_{\rm ph}$, or a full redshift posterior $p(z)$
\citep[e.g.][]{leistedt16a}. These point estimates are used to 
divide galaxies into tomographic bins. We define \nph{z} as the 
distribution of photo-$z$ point-estimates.

In principle surveys can use the calibration sample to parameterise 
$p(z_{\rm ph}|z;p_{\mu})$, the probability distribution of measuring 
$z_{\rm ph}$ for a galaxy at true (spectroscopic) redshift $z$; $p_\mu$ 
are parameters describing the distribution. Using 
$p(z_{\rm ph}|z;p_{\mu})$ and \nph{z}, one can obtain the underlying 
true galaxy distribution \ntrue{z} from
\begin{equation}
  \label{eqn:zdis}
  \nph{z_{\rm ph}}=\int \ntrue{z}p(z_{\rm ph}|z;p_\mu) dz
\end{equation}
\citep{ma08}. Using this information, one can also infer the true 
redshift distribution in the $i$th tomographic bin, \nttrue{z}. 
We define $\nest{z}$ and \ntest{z} as estimators for \ntrue{z} and 
\nttrue{z}.

In practice, surveys only use photo-$z$ point estimates to assign 
galaxies to tomographic bins using various methods. For example, 
some surveys compute \ntest{z} by stacking the redshift posteriors 
$p(z)$ of all galaxies in a given bin \citep{kilbinger13a,troxel17a}. 
Others have computed \ntest{z} by dividing samples of galaxies with 
known redshifts into tomographic bins and weighting the resulting 
redshift distributions based on the photometric properties of the 
surveyed galaxies \citep{hildebrandt17a,hikage18a}. Different methods 
of estimating \ntrue{z}\ can bias constraints \cite[e.g.][]{hikage18a}.
For simplicity, in our simulated survey, we estimate \nttrue{z}\ with 
\ntph{z}, the distribution of photo-$z$ point-estimates in a given 
bin.

An estimator \estp\ of the true cosmological parameters is then
obtained via a comparison of the expectation values of a summary
statistic $\Bd(\Bp)$ and its observed values \dobs. Because $\Bd(\Bp)$
and \dobs\  respectively depend on \nest{z} and \ntrue{z}, inaccuracies
in \nest{z} will bias \estp. Our analysis focuses on quantifying the
effects of such inaccuracies arising from two classes of photo-$z$
errors: (1) residual photo-$z$ bias and (2) directly approximating
\ntrue{z} with some variation of \nph{z}.

We note that instead of using $z_{\rm ph}$ for ray-tracing, a real survey might
instead randomly draw a galaxy's redshift for each realisation from
its $p(z)$ \citep{liu15a}, or from the estimated true galaxy redshift
distribution in its tomographic bin \citep[analogous to the procedure
in][]{kacprzak16a}. Both alternatives mitigate the impact of inaccuracies 
in \nest{z} on \estp\ at the cost of slightly weaker constraints.

\subsubsection{Residual photo-$z$ Biases}\label{section:m_zph_bias}

Uncalibrated photo-$z$ error refers to the errors that propagate to the 
inferred \estp\ from the uncertainties in the parameters $p_\mu$ in 
$p(z_{\rm ph}|z;p_{\mu})$. Such uncertainties can occur, for example, due 
to the finite size of the spectroscopic calibration sample
\citep{huterer06a, ma08}.  Because photo-$z$ bias $b_{\rm ph}=z_{\rm ph}-z$
has the leading order effect on biases in \estp\ for
$P^{\kappa\kappa}$ \citep{ma06, huterer06a}, we only account for the
impact of bias in our assessment of uncalibrated errors; we assume
that there is no scatter and catastrophic error. The photo-$z$
requirements listed in the LSST science book \citep{lsst09a} demand
$|b_{\rm ph}|<0.003(1+z)$.

Under the assumptions of no scatter and catastrophic error,
the impact of calibrated component of $b_{\rm ph}$, $b_{ph,cal}$, on \estp\ can 
be entirely removed by setting $\hat{n}(z)$ equal to $\nph{z+b_{\rm ph,cal}}$.
Therefore, we further assume that the uncalibrated $b_{\rm ph}$
(i.e., a residual bias left after some calibration procedure is performed)
is the only source of error in our assessment. Hereafter, we refer 
to this as residual photo-$z$ bias. Consequently, 
$\ntrue{z} = \nest{z+b_{\rm ph}}$ and its best available
estimate after a hypothetical calibration procedure is performed is
$\nest{z} = \nph{z}$. To assess the impact of different levels of residual 
bias, we vary \ntrue{z}.

We further subdivide our analysis of $b_{\rm ph}$ into two
subcases. To build our intuition, we first employ
the Single-$z$ dataset to treat the simple case where all of the
galaxies are distributed across the sky at a single $z$ (and so do not
include tomography). For the second case, we use the Tomographic
dataset to address the impact of residual $b_{\rm ph}(z)$ for more
realistic galaxy distributions. In this case, we parameterise $b_{\rm ph}$
with $b_{\rm ph}(z)=b(1+z)$ where $b$ is a constant;
a more realistic analysis would allow $b$ to vary with $z$. Unlike
\citetalias{petri16b} who effectively assumed $b= 0.003$, we
investigate the biases in \estp\ arising from different values of $b$.

\subsubsection{Realistic photo-$z$ errors}
\label{section:m_realistic_error}

\begin{figure}
  \center
\includegraphics[width = 3.5 in]{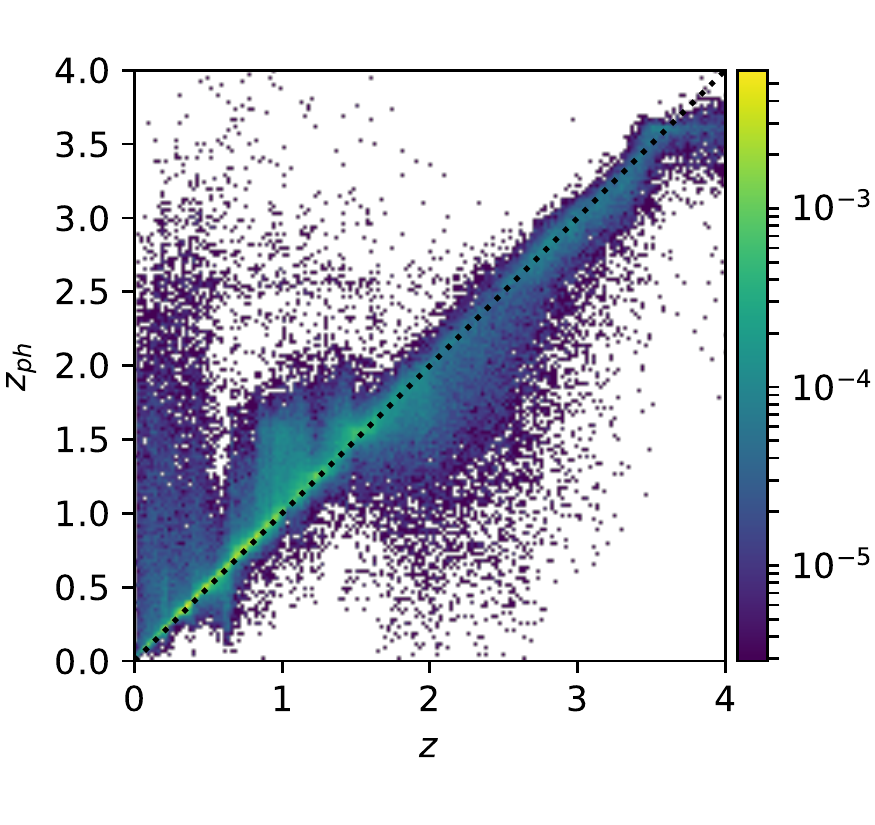}
\caption{\label{figure:photoz_pdf} 2D histogram illustrating the simulated 
  photo-$z$ performance for LSST adopted from \citet{rhodes17a}. The photo-$z$ 
  values are the expectation values of the redshift PDFs $p(z)$ calculated for 
  each galaxy in the COSMOS catalogue \citep{laigle16a}, using spectral templates 
  and photometry for filters that LSST will use. We refer to the PDF of measuring 
  a photo-$z$ point estimate of a galaxy at $z$, $p(z_{\rm ph}|z)$, as the photo-$z$ 
  PDF. To generate this figure, we omit all galaxies with $z>4$ or 
  $z_{\rm ph}>4$, define bins with $\Delta z=\Delta z_{\rm ph}\sim0.15$ and colour the bins 
  by the fraction of galaxies that lie within the bin. The black dotted line 
  indicates the 1-to-1 line along which photo-$z$ measurements have no errors.}
\end{figure}

The second class of errors we explore are motivated by simplified techniques 
to compute \nest{z}. The simplest approximation of \ntrue{z} is setting 
$\nest{z}=\nph{z}$, without making any modifications to 
\nph{z} to account for calibrated photo-$z$ errors. We will refer to errors 
from this approximation as unmodelled realistic errors since the analysis folds in 
the realistic $p(z)$, but does not attempt to reduce the error.

As surveys become more sensitive, better approximations of \ntrue{z}
become necessary. Recent analyses of the DES Science Verification data 
\citep[SV;][]{abbott16a}, the DES Year 1 data \citep[Y1;][]{troxel17a}, and 
the HSC first-year data \citep{hikage18a} made leading order corrections to 
their respective estimates of \ntrue{z}; they remove the centroid biases 
$\delta z^i$ from their estimates of \nttrue{z}. We define centroid bias as 
$\delta z^i = \langle z_{\rm est}^i\rangle - \langle z^i\rangle$, i.e. the 
difference between the means of \ntest{z} and \nttrue{z}. They then 
approximate \nttrue{z} by a shifted version 
$\nttrue{z} \approx \ntest{z-\delta z^i}$. After the centroid shift, the 
remaining errors include residual errors for $\delta z^i$  \citep{huterer06a} 
and errors in the shape of $\hat{n}^i(z)$. Though the latter effect does not 
significantly bias the DES Y1 results \citep{troxel17a}, it is
more important for the HSC first-year results \citep{hikage18a}.

For realistic LSST photo-$z$ performance, we assess the impact of 
unmodelled realistic photo-$z$ errors and the uncertainties in 
the shape of $\hat{n}^i(z)$ on inferred cosmological parameters. For the 
latter case, we approximate \nttrue{z} with \ntph{z} after removing centroid 
bias. We use the Tomographic dataset and model photo-$z$ performance with 
a recently simulated spectroscopic calibration sample, 
employed by \citet{rhodes17a}. Figure~\ref{figure:photoz_pdf} illustrates 
this dataset. To capture the photo-$z$ performance of the LSST filter set, 
the dataset was constructed by applying a basic template-based method to 
the COSMOS catalogue \citep{laigle16a}. However, a variety of factors are not modelled,
such as the dependence on depth, data, quality, or selection function 
\citep{rhodes17a}.

As we discuss in detail in Appendix~\ref{app:photoz_perf}, these photometric
redshifts do not meet the LSST photo-$z$ requirements. Despite
these drawbacks, it gives an estimate of LSST's photo-$z$ PDF (i.e. $p(z_{\rm ph}|z)$, 
the probability density function for measuring a photo-$z$ value at a given 
spectroscopic redshift), and will suffice for assessing the relative
resilience of $n_{\rm pk}$ and $P^{\kappa\kappa}$ to these types of photo-$z$ errors.

To simulate photo-$z$ errors, we divide the simulated spectroscopic
calibration sample into 40 uniform bins spanning $0<z_{\rm ph}<3$; each
bin contains at least 987 galaxies. Within each photo-$z$ bin, we
compute normalised histograms of $e_z\equiv \frac{z_{\rm ph}-z}{1+z}$ with
the number of bins given by the Freedman--Diaconis Estimator
\citep{freedman81a}. For simplicity, we assume that these histograms
perfectly describe the distribution of photo-$z$ errors and that
$p\left(e_z|z_{\rm ph}\right)$ is the same for both the simulated
calibration set and our simulated survey. 

To assess the impact of unmodelled realistic errors we randomly
draw the true source redshift $z$ for each galaxy in \scobs\ from
$p\left(e_z|z_{\rm ph}\right)$
when we set $z_{\rm ph}$
to $z_g$.  We then perform ray-tracing simulations to produce new
\scobs. Each realisation of \scobs\ uses a different random seed to
draw true source redshifts. All randomly drawn values falling outside
of $0\leq z\leq4.1925$ are set to the closest value in the range;
$z=4.1925$ is slightly smaller than the redshift of the furthest lens
plane in our Tomographic dataset. Figure~\ref{figure:realistic_n(z)} below shows a comparison
between \nph{z} and the average \ntrue{z}.

To assess the impact of errors in the shape of $\hat{n}^i(z)$, we follow 
the same procedure outlined in the preceding paragraph with a modification. 
After producing the source redshifts for a given realisation, 
we remove the centroid bias between \nttrue{z}\ and \ntph{z}\ from the 
source redshifts of the galaxies in bin $i$ before performing ray-tracing. 
In other words, we modify \nttrue{z} to be equal to \nttrue{z+\delta z^i}. 
This approach is equivalent to assuming that a calibration process was
able to obtain the correct centroid of the redshift-distribution in
each tomographic bin.  We believe this is a reasonable approximation
for setting $\hat{n}^i(z) = n^i_{\rm ph}(z-\delta z^i)$; in any case it
allows us to isolate errors arising purely from differences in the 
shapes of \nttrue{z} and $\hat{n}^i(z)$.

\begin{figure}
  \center
\includegraphics[width = 3.5 in]{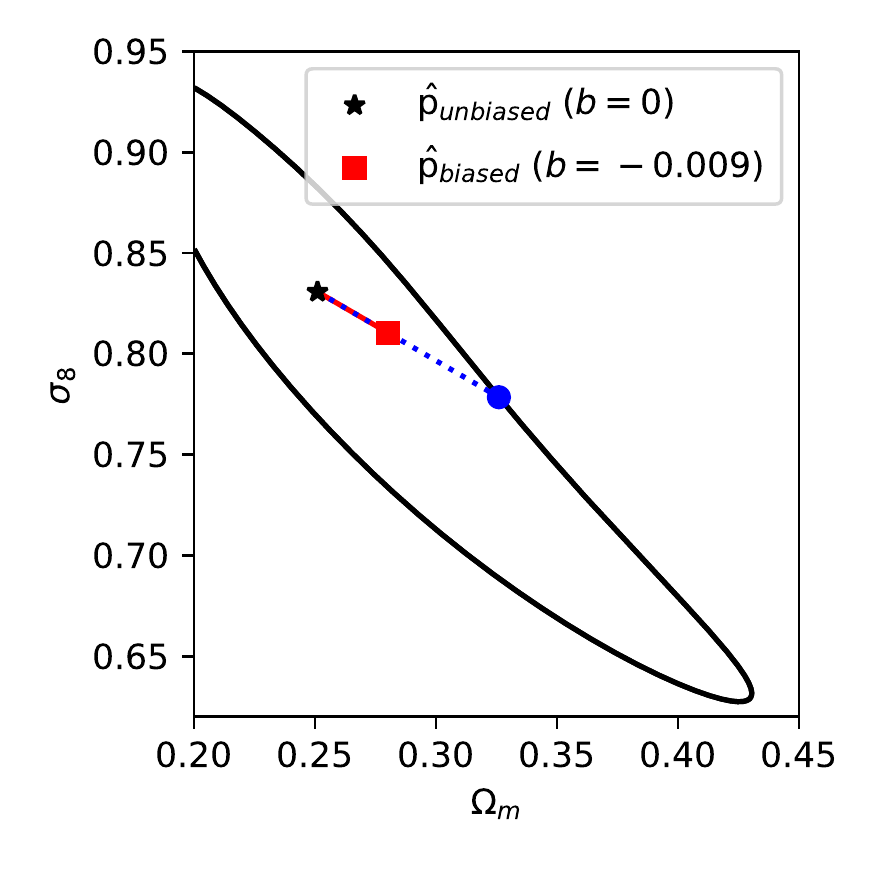}
\caption{\label{figure:relbias} Illustration of the calculation of the 
  figure-of-merit for the parameter bias, \relbias. The black star and 
  contour indicate the most likely $\Lambda$CDM cosmology 
  ($\estp_{unbiased}$) and $68\%$ confidence contour inferred from the 
  tomographic $n_{\rm pk}(\nu)$ without photo-$z$ errors. The red star indicates 
  the most-likely value inferred with photo-$z$ bias, given by 
  $b_{\rm ph}(z)=b(1+z)$ with $b=-0.009$. \relbias\ is the ratio 
  between the lengths of the solid red line (the magnitude of parameter bias 
  and the blue dotted line (the distance from $\estp_{unbiased}$ to the unbiased
  $68\%$ confidence contour along the direction of parameter bias).
  }
\end{figure}

\subsection{Cosmological Parameter Inference}\label{section:inference}
\begin{table}
  \centering
  \caption{ Brief summary of cosmological
    models and the maximum survey size (above which we can no longer
    reliably measure the parameter biases \relbias; see \S~\ref{section:b/u})
    for the two summary statistics
    we employ: peak counts ($n_{\rm pk}$) and tomographic power
    spectrum ($P^{\kappa\kappa}$).  For the latter, we employ a
    principal component analysis (PCA) and keep $N_c=30$ components (see \S~\ref{section:PCA}). }
  \label{tab:cosmo_models}
  \begin{tabular}{lcccc}
    \hline
    model & ${\bf p}$ & Points in $\chi^2$ grid & \multicolumn{2}{c}{$\Omega_{max}$ ($\rm{deg}^2$)} \\
    &&& $n_{\rm pk}$ & $P^{\kappa\kappa}(N_c=30)$
    \\
    \hline
    $\Lambda$CDM & $(\Omega_m,\sigma_8)$     & $1000^2$ & 984  & 15033 \\
    $w$CDM       & $(\Omega_m, w, \sigma_8)$ & $100^3$  & 2002 & 24930 \\
    \hline
  \end{tabular}
\end{table}

Bayes' theorem allows us to synthesise the posterior probability
distribution of \estp, assuming some model $M$, from an
observed summary statistic vector
$\hat{\bf d}^{obs}$ and prior knowledge $p({\bf p}|M)$ using
\begin{equation}
  \label{eqn:bayes}
  p({\bf p}|\hat{\bf d}^{obs},M) = \frac{p(\hat{\bf d}^{obs}|{\bf p},M)
    p({\bf p}|M)}{p(\hat{\bf d}^{obs}|M)}.
\end{equation}
In this equation, $p(\hat{\bf d}^{obs}|M)$ gives the probability
of observing $\hat{\bf d}^{obs}$ for any cosmology while $p(\hat{\bf
  d}^{obs}|{\bf p},M)$ is the likelihood function for observing
$\hat{\bf d}^{obs}$ in cosmology ${\bf p}$. We infer the posterior
of \estp\ using two separate cosmological models: $\Lambda$CDM and
$w$CDM. Tables~\ref{tab:sim_overview}~and~\ref{tab:cosmo_models} list
details about inference in each model.

 For our analysis, we treat $p({\bf p}|M)$/$p(\hat{\bf d}^{obs}|M)$ as
 a normalisation constant within the region sampled by \pset\ and zero
 elsewhere. We assume that our summary statistics follow a
 multivariate Gaussian distribution with covariance matrix
 ${\bf{C}}$. Thus, ${\rm log}\ p({\bf p}|\hat{\bf d}^{obs},M) \propto
 -\chi^2(\hat{\bf d}^{obs}|{\bf d}({\bf p}), {\bf C})$, where
\begin{equation}
  \chi^2(\hat{\bf d}^{obs}|{\bf d}({\bf p}), {{\bf C}}) = 
  (\hat{\bf d}^{obs} - {\bf d}({\bf p}))^T {{\bf C}^{-1}}
  (\hat{\bf d}^{obs} - {\bf d}({\bf p})).
\end{equation}
We assume that ${\bf C}$ is independent of cosmology, and estimate it
using the $N_r$ realisations of the summary statistics $\hat{\bf d}_r$
computed for ${\bf p_0}$ with \kfid\ or \ktfid. The formula for the
unbiased estimate of ${\bf C}$ is
\begin{equation}
  \hat{\bf C} = \frac{1}{N_r - 1}\sum_{r=1}^{N_r}
  (\hat{\bf d}_r - \Bd(\pfid))(\hat{\bf d}_r - \Bd(\pfid))^T.
\end{equation}
To make the inverse of $\hat{\bf C}$ an unbiased estimator of
${\bf C}^{-1}$, we rescale it with
\begin{equation}
  \hat{{\bf C}^{-1}} = \frac{N_r - N_b -2}{N_r-1}{\bf C}^{-1},
\end{equation}
where $N_b$ is the number of components of the summary
statistic \citep{hartlap07a}.

To evaluate the likelihood function for a \Bp\ not included in \pset,
we make use of cosmology interpolators. We follow
\citetalias{zorrilla16a} for the Single-$z$ dataset and interpolate
$\chi^2(\hat{\bf d}^{obs}|{\bf d}({\bf p}), \hat{{\bf C}})$
at an arbitrary cosmology \Bp\ using the values at each cosmology in \pset.
However, unlike \citetalias{zorrilla16a}, we use cubic interpolation
rather than linear interpolation, as the latter will {\it always}
identify a cosmology in \pset\ as the most-likely cosmology \estp. 
For the Tomographic dataset, we instead employ an emulator which
interpolates ${\bf d}({\bf p})$ between each cosmology in \pset\ using
augmented RBF interpolation with a cubic basis function
\citep{petri16c}. Our choice of interpolation differs from that of
\citetalias{petri16b} and we refer the reader to
Appendix~\ref{app:interp_comp} for a comparison of various
interpolation schemes. 

Since we only infer $2-3$ cosmological parameters, we perform inference by 
sampling a grid using the interpolator. Hereafter we refer to that grid as 
the sampled $\chi^2$ grid. Tables~\ref{tab:sim_overview}~and~\ref{tab:cosmo_models}
provides details about the sampling of these grids.

Our assessment of the impact of unmodelled photo-$z$ errors 
and of the errors in the shape of $n^i(z)$ requires a different approach. 
In both cases, \ntrue{z} is randomly drawn for each realisation, yielding 
different values of \Bd. Consequently, we adopt the basic procedure of 
constructing separate grids of $\chi^2$ values for each realisation. Then 
for each point on the grid, we average the $\chi^2$ values over all $N_{r}$ realisations.

\subsubsection{Quantifying Inferred Parameter Biases}\label{section:b/u}

Any discrepancies between $\nest{z}$ and $\ntrue{z}$ will lead to biases
on \estp. We can debias the results by propagating the
uncertainties in $\hat{n}(z)$ to \estp\ \citep[e.g.][]{huterer06a,abbott16a}.
Unfortunately, this is computationally prohibitive for non-analytic
statistics, like peak counts because it requires rebuilding the
interpolator (and all of the ray-tracing simulations) for every
sampled $\hat{n}(z)$. To circumvent this cost in their analysis
of DES SV data using scaled peaks counts, \citet{kacprzak16a} parameterised
the linear impact that the centroid bias of the entire galaxy distribution
(they didn't use tomography) had on the heights of the scaled peaks. For our 
analysis, we choose not to directly propagate the uncertainty.

Instead, we define a bias-to-uncertainty ratio \relbias, as
the ratio of the magnitude of the bias $|\Delta {\bf p}|\equiv |\estp_{biased} - \estp_{unbiased}|$
and the distance between the unbiased most-likely \estp\ and the unbiased
$68\%$ confidence contour along the direction of the bias $|\Delta {\bf p}|$.
Figure~\ref{figure:relbias} presents a graphical illustration of \relbias\ 
for tomographic $n_{\rm pk}(\nu)$ in a case with uncorrected photo-$z$ bias.
The above definition is necessary, because the posteriors can change their shapes due to discrepancies 
between $\nest{z}$ and $\ntrue{z}$. \relbias\ is defined such that its 
value is independent of the relative normalisation of the different components of \Bp.

We employ \relbias\ as a proxy for the degradation in the constraints
since the degraded posterior needs to include both the biased and
unbiased most likely values of \estp.  As we scale our survey
to larger sizes, and the posterior more closely resembles a symmetric
Gaussian, \relbias\ becomes a better proxy for posterior
degradation. We identify the case where $\relbias = 1.5$ as the
benchmark for when photo-$z$ errors contribute considerable uncertainty
to the inferred \estp. This value corresponds to an error degradation
of $\sim 50\%$, which is analogous to the benchmark employed by
\citet{huterer06a} while measuring degradations in marginalised
uncertainties.

\subsubsection{Scaling Survey Size} \label{section:scaled_size}

To directly compare our results from the Tomographic
dataset to a large survey such as LSST, we need to scale our
$\chi^2$ values to be consistent with a survey subdivided into
$N$ $(3.36{\rm deg})^2$ subfields. To do this we
assume that our subfield is an average subfield in the survey, and we
multiply all of the $\chi^2$ values in the sampled grid by a factor
of $N$.

The finite number of cosmologies in \pset\ introduces error to the
interpolated $\Bd(\Bp)$ that will propagate to errors in the computed
$\chi^2$ value. For each summary statistic, we compute the average
magnitude of this propagated interpolation error, by
identifying the 10 closest values in \pset\ to the most likely
\estp\ and employing
\begin{equation}
  \frac{1}{10}\sum_{j=1}^{10} \left|
  \frac{\chi^2(\hat{\bf d}^{obs}|{\bf d}({\bf p}_j), \hat{{\bf C}})
    - \chi^2(\hat{\bf d}^{obs}|\hat{\bf d}({\bf p}_j), \hat{{\bf C}})}
       {\chi^2(\hat{\bf d}^{obs}|{\bf d}({\bf p}_j), \hat{{\bf C}})}
       \right|,
\end{equation}
where $\hat{\bf d}({\bf p}_j)$ is emulated from \pset\ excluding $\Bp_j$.

This interpolation error corresponds to a maximum survey size for
which we can reliably compute \relbias.  This is because as the survey
is scaled up, the $\chi^2$ values increase, the $68\%$ confidence contour
shrinks, and it eventually corresponds to a small region that is buried
in the numerical noise in the $\chi^2$ surface. Unless the
difference between $\chi^2_{\rm min}$ and the $\chi^2$ value corresponding to 
the $68\%$ confidence contour is larger than the numerical interpolation error, 
we can not reliably measure the distance between the most-likely \estp\ and
the contour.  Table~\ref{tab:cosmo_models} lists this maximum survey
size for each cosmological model and summary statistic.

In addition to numerical limits on computing the bias \relbias,
the original $\chi^2$ grid can also become too poorly sampled 
to reliably infer the posterior for large $N$ in the $w$CDM cosmology.
These cases are identified when the marginalised standard deviation of
a component \estp\ is smaller than the sampling resolution in that
component. When these cases arise, we simply resample the grid with
$101^3$ grid points centred on the original most-likely \estp\ and
spaced with resolutions at least half the size of the marginalised standard
deviations. Because of the slightly higher resolution grid,
the most likely-value \estp\ may shift. For consistency, 
we always compute the \relbias\ for a given statistic in a $w$CDM
cosmology with respect to the $\estp$ with the lowest unscaled $\chi^2$
value encountered on any grid.

Our assessment of biases in \estp\ that arise from errors in the shapes of 
$n^i(z)$ involves a slightly modified procedure. To construct a 
survey of $N$ subfields, we group the $N_r=1000$ realisations into as 
many independent groups of $N$ realisations as possible without having 
any realisations repeat groups and discard all remaining realisations not 
assigned to a group. Then when adjusting the randomly drawn values of 
\nttrue{z} before ray-tracing, we use a single correction value for 
all realisations in a given group that removes the centroid bias
between the group's combined \ntph{z} and \nttrue{z}. 
For a survey of $N>1$ subfields, this allows the 
\ntph{z} in individual subfields to still have centroid bias,
which then cancel out when these individual subfields are aggregated.
Consequently, our assessment of the impact of errors in the shape of 
$n^i(z)$ in surveys of size $N$ effectively includes ${\rm floor}(1000/N)$ 
unique realisations of the survey produced from 
${\rm floor}(1000/N)\times N$ realisations of \scobs.

\subsubsection{Principal Component Analysis} \label{section:PCA}

As in \citetalias{petri16b}, we attempt to apply Principal
Component Analysis (PCA) to reduce the dimensionality of the summary
statistics in the Tomographic dataset. This is necessary
because, as mentioned above, the full power spectrum has 825
components, which is $\sim 5\%$ of the $N_r=16,000$ realisations
used to construct the emulator and covariance matrix.
As a first step, we compute the 
mean $\langle {\bf d} \rangle$ and the variance $\sigma_{\bf d}$
of each statistic, over all cosmologies in \pset. 
We then follow the procedure in \citetalias{petri16b},
except for one difference.  While the PCA in  \citetalias{petri16b}
is performed on the normalised components  $[d_i - \langle {\bf d}_i \rangle]/\langle {\bf d}_i \rangle$,
we instead use the whitened components $[d_i - \langle {\bf d}_i \rangle]/\sigma({\bf d_i})$.
Consequently, in our case, each component has unit variance before applying PCA 
\citep{ivezic14a}, although they still have co-variance.

We refer the reader to Appendix~\ref{app:PCA} for a detailed
explanation for our choices of the number of principal components
$N_c$ for each statistic. We ultimately choose not to use PCA for the
$n_{\rm pk}$ (equivalent to $N_c=100$) and to use $N_c=30$ for
$P^{\kappa\kappa}$. Table~\ref{tab:cosmo_models} lists the maximum
survey sizes for which we can compute \relbias\ using $P^{\kappa\kappa}$
with $N_c=30$ components and the peak counts using the full set of components.

\begin{figure*}
  \center
\includegraphics[width = 6.9 in]{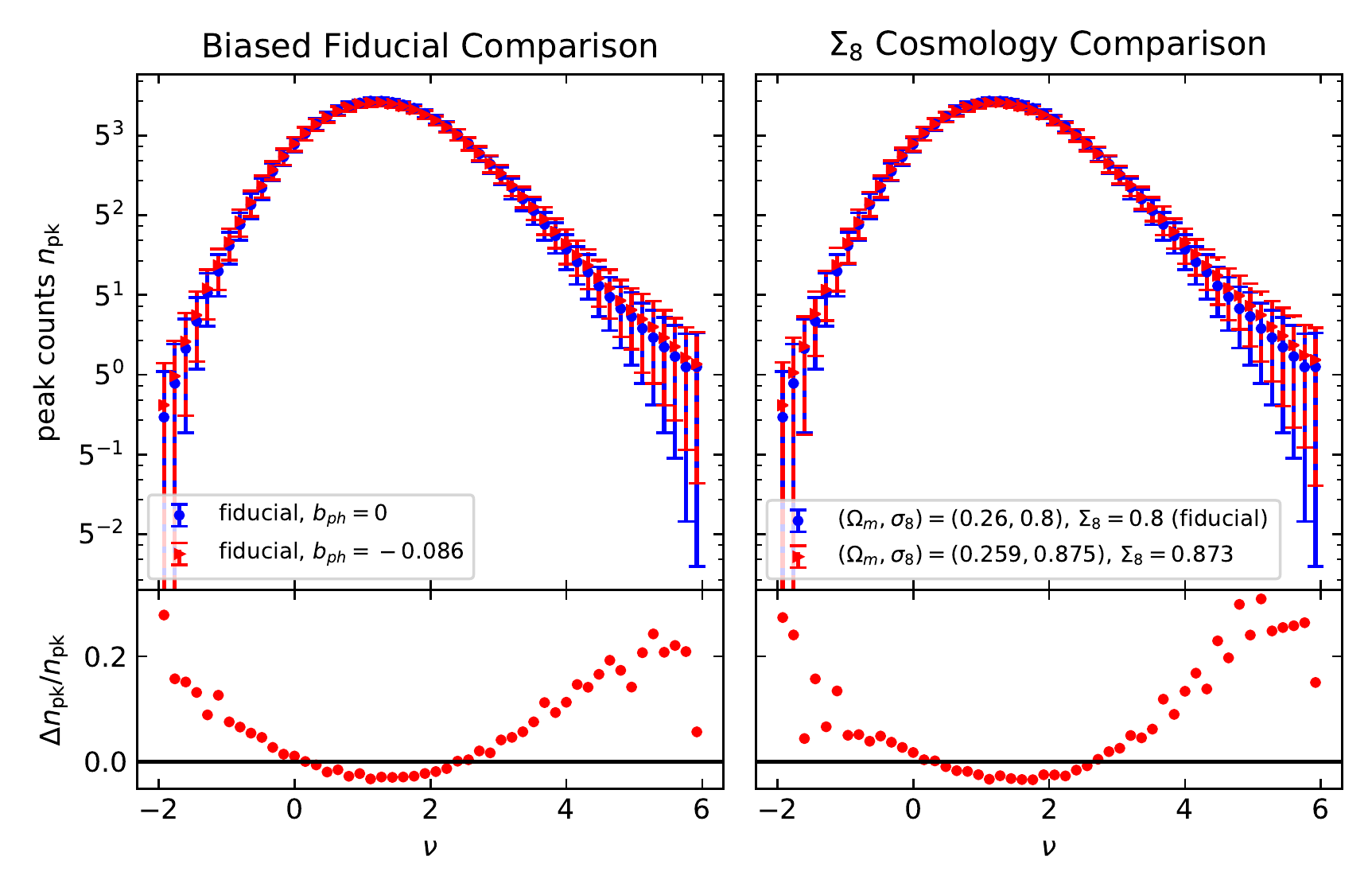}
\caption{\label{figure:simple_bias_hist_comp} Comparisons between
  the average peak count histograms in the fiducial cosmology with
  and without a constant photo-$z$ bias $b_{\rm ph}$ (left panel),
  and of a different cosmology
  $(\Omega_m,\sigma_8) = (0.259,0.875)$ with $b_{\rm ph}=0$ (right
  panel). The cosmology in the right comparison was selected because
  it has a similar $\Sigma_8$ value to the most likely $\Sigma_8$ from
  the average peak histogram in the fiducial cosmology with a bias of
  $b_{\rm ph}=-0.086$. This side-by-side comparison qualitatively
  illustrates how $b_{\rm ph}$, in our simple case, creates 
  deformations in the peak count histogram that can be mimicked
  by cosmologies with a higher $\Sigma_8$. (Note that we use 50 bins 
  in this figure for clarify of presentation; in the calculations we use a finer
  binning with 100 bins.)
  }
\end{figure*}

\begin{table}
  \centering
  \caption{ The table shows the biased
    inferred best-fit $\Sigma_8$ values for source planes at various
    biased redshifts.  We also list the reduced $\chi^2$ value for the
    best-fit cosmology using the average peak histogram.}
  \label{tab:simple_S8}
  \begin{tabular}{cccccc}
    \hline
    $b_{\rm ph}$ & $\Sigma_8^{(a)}$ & reduced $\chi^2{}^{(b)}$ &
    $\left<\Sigma_8^{r}\right>$ & median$\{\Sigma_8^{r}\}$ 
    & std$\{\Sigma_8^{r}\}$ \\
    \hline
    $-0.233$ & 0.946 & $+0.00352$ & 0.929 & 0.926 & 0.0421 \\
    $-0.183$ & 0.912 & $+0.00282$ & 0.906 & 0.905 & 0.0402 \\
    $-0.134$ & 0.893 & $+0.00221$ & 0.880 & 0.880 & 0.0397 \\
    $-0.086$ & 0.866 & $+0.00279$ & 0.856 & 0.857 & 0.0397 \\
    $-0.040$ & 0.832 & $-0.00006$ & 0.831 & 0.829 & 0.0362 \\
    0 & 0.8   & 0 & 0.809 & 0.799 & 0.0296 \\
    $+0.005$ & 0.797 & $-0.00385$ & 0.808 & 0.798 & 0.0306 \\
    $+0.049$ & 0.787 & $+0.00263$ & 0.787 & 0.787 & 0.0334 \\
    $+0.092$ & 0.766 & $-0.00092$ & 0.764 & 0.767 & 0.0371 \\
    $+0.135$ & 0.756 & $+0.00026$ & 0.740 & 0.745 & 0.0399 \\
    $+0.176$ & 0.706 & $-0.00303$ & 0.715 & 0.714 & 0.0413 \\
    $+0.216$ & 0.683 & $-0.00443$ & 0.686 & 0.693 & 0.0454 \\
    \hline
  \end{tabular}
  \\$^a$ The most likely $\Sigma_8$ for the average peak histogram
  \\$^b$ Negative values are an artifact of the interpolation of $\chi^2$ values.
\end{table}

\begin{figure}
  \center
\includegraphics[width = 3.5 in]{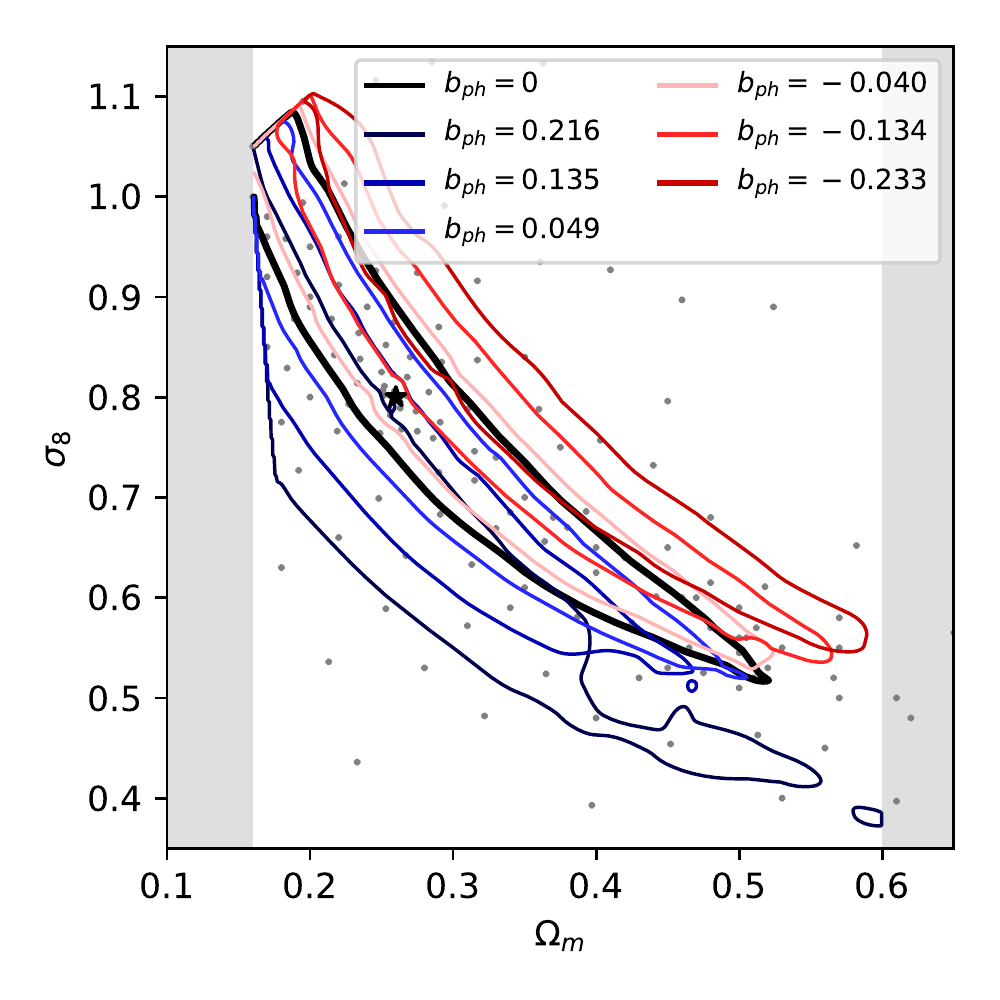}
\caption{\label{figure:shifted_contours} $95.5\%$ parameter confidence
  contours inferred for the fiducial cosmology when the source plane
  is placed at redshifts $z$ biased by various constant values of
  $b_{\rm ph}$. The black star indicates the true cosmology.
  As this plot illustrates, the impact of a constant photo-$z$ bias is to
  shift the contours perpendicular to the degenerate direction.}
\end{figure}

\begin{figure}
  \center
\includegraphics[width = 3.5 in]{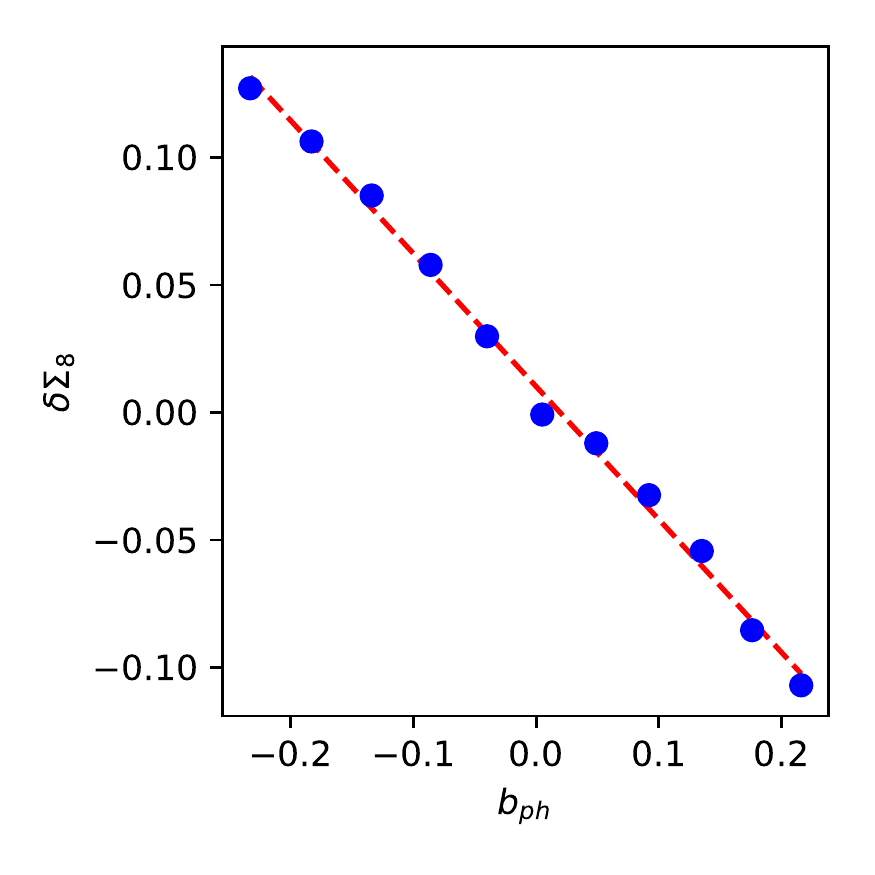}
\caption{\label{figure:simple_S8_bias} The parameter bias
  ($\delta\Sigma_8$) as a function of constant photo-$z$ bias
  ($b_{\rm ph}$). We compute $\delta\Sigma_8$ using the
  median($\Sigma_8^{r}$) of the best fits over all realisations of the
  mock observations.  The error-bars, obtained from
  std$(\Sigma_8^{r})$, are smaller than the markers. The red dashed
  line illustrates the least squares error-weighted best fit line, given
  by $\delta\Sigma_8 = (0.5211\pm0.0002) b_{\rm ph} + (0.01020\pm 3\times10^{-5})$}
\end{figure}

\begin{figure*}
  \center
\includegraphics[width = 5in]{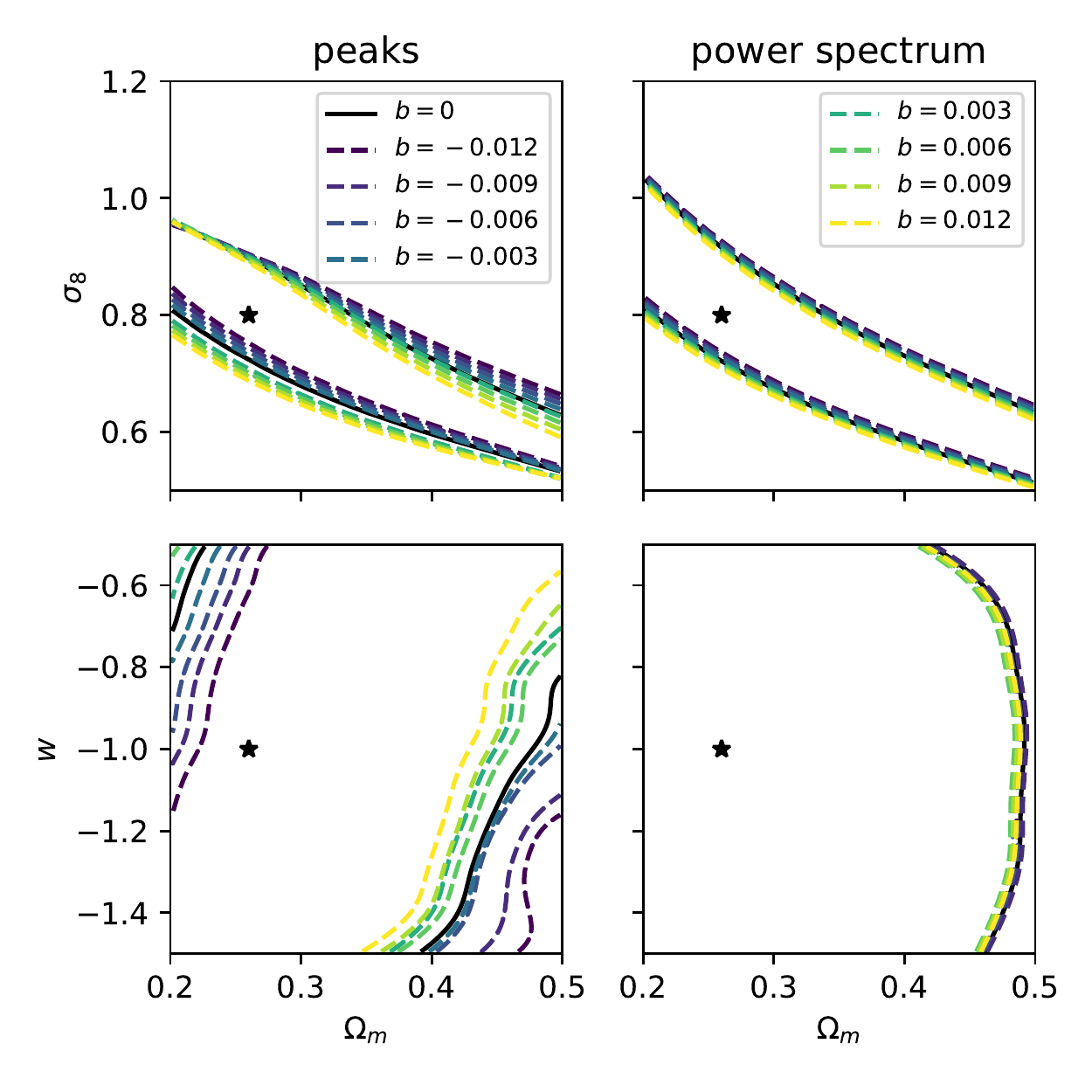}
\caption{\label{figure:small_const_bias_contours} The $95.5\%$
  confidence contours for the tomographic $n_{\rm pk}(\nu)$ (left
  column) and $P^{\kappa\kappa}(l)$ (right column) for different small
  biases parameterised by $b$ in a $w$CDM cosmology. Each
  panel in the top (bottom) row shows the contours for the posteriors
  marginalised over $w$ ($\sigma_8$). The black star shows the
  fiducial cosmology used to generate the observations.}
\end{figure*}

\begin{figure*}
  \center
\includegraphics[width = 5in]{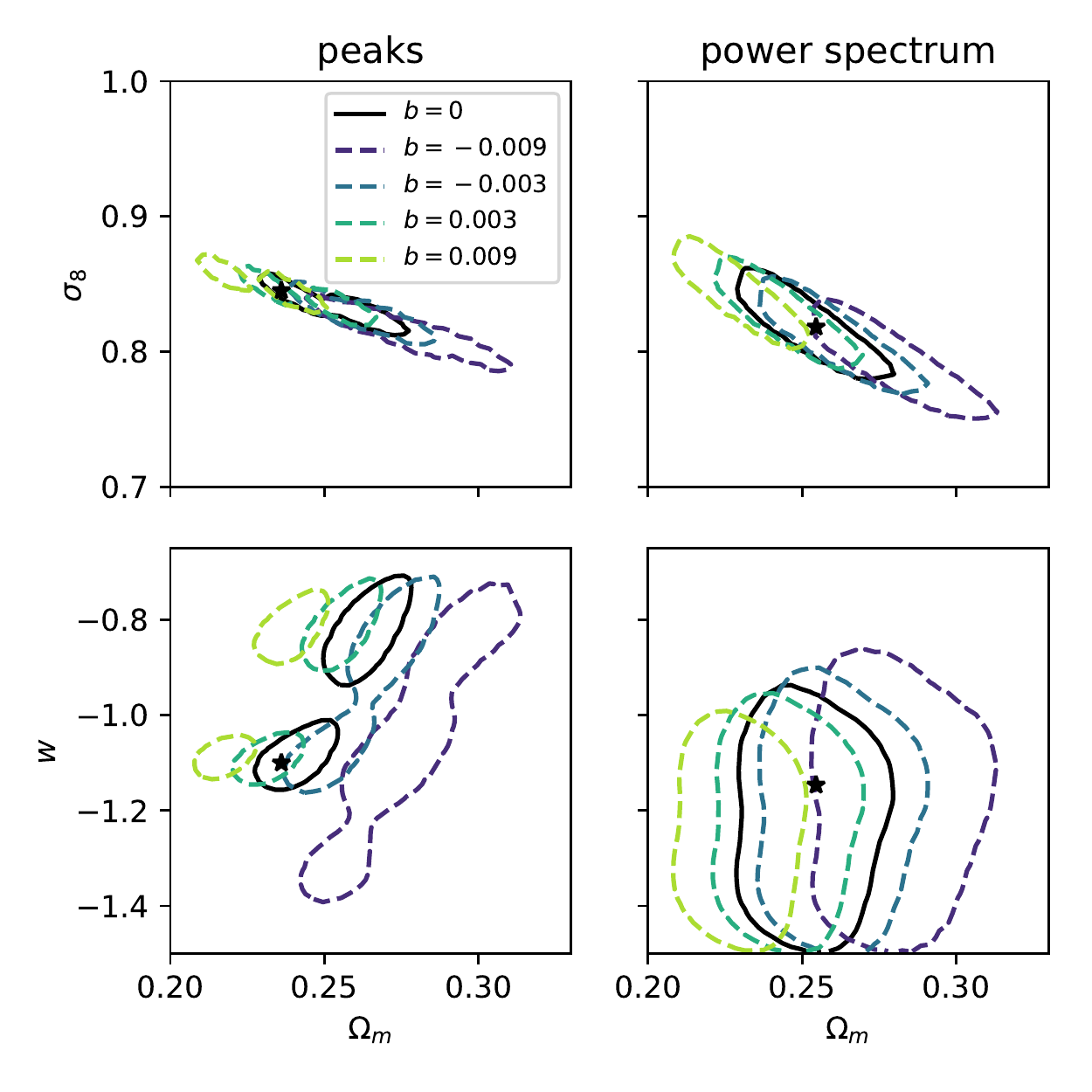}
\caption{\label{figure:small_const_bias_contours_N177} Same as
  Fig.~\ref{figure:small_const_bias_contours}, except that we have
  scaled the likelihood values to those of a survey made up of $N=177$
  subfields ($\sim 2000~{\rm deg}^2$). In this figure, stars mark the
  most-likely values inferred either from peak counts or from the
  power spectrum.}
\vspace{\baselineskip}
\end{figure*}

\begin{figure*}
  \center
\includegraphics[width = 5in]{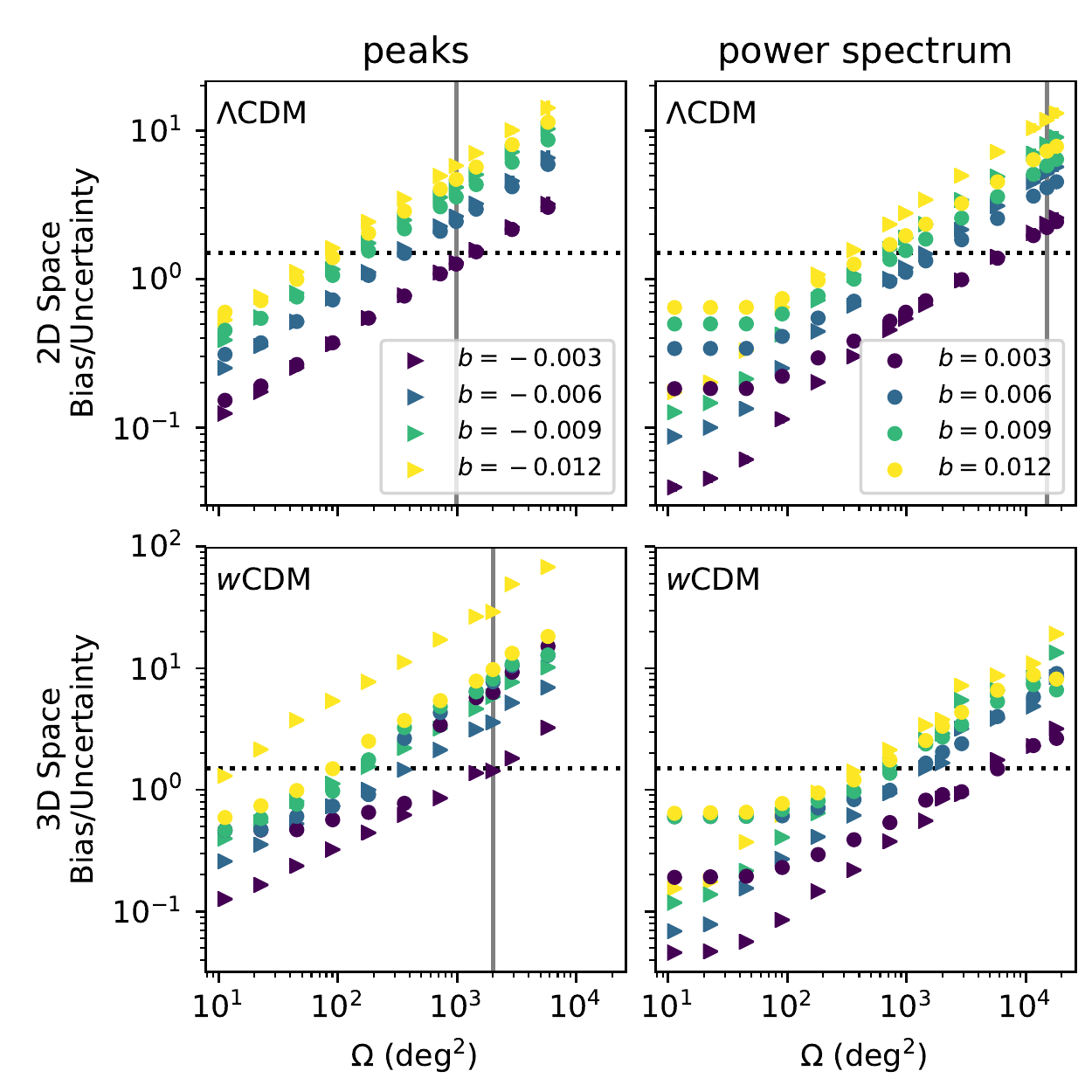}
\caption{\label{figure:bias/uncertainty}
  Figure of merit \relbias\ for the bias resulting from photo-$z$ errors
  parameterised by constant values of bias parameter $b$.
  \relbias\ is the size of the bias in the inferred cosmological
  parameters, relative to the distance between the unbiased most
  likely value to the $68\%$ confidence contour, along the direction
  of the bias. The horizontal dashed line indicates where this ratio
  has a value of 1.5 (our proxy for degrading the constraints by
  $~50\%$). The top (bottom) row shows \relbias\ in a $\Lambda$CDM
  ($w$CDM) cosmology, and the left (right columns) are for peaks
  (power spectrum). The colors and shapes of the markers correspond to
  different bias parameters $b$ as labelled. The vertical gray line
  marks the maximum scaled up survey size for which our results can be
  trusted (points to the right of this line may have
  significant interpolation errors).
  }
\vspace{\baselineskip}
\end{figure*}

\begin{figure}
  \center
\includegraphics[width = 3.5 in]{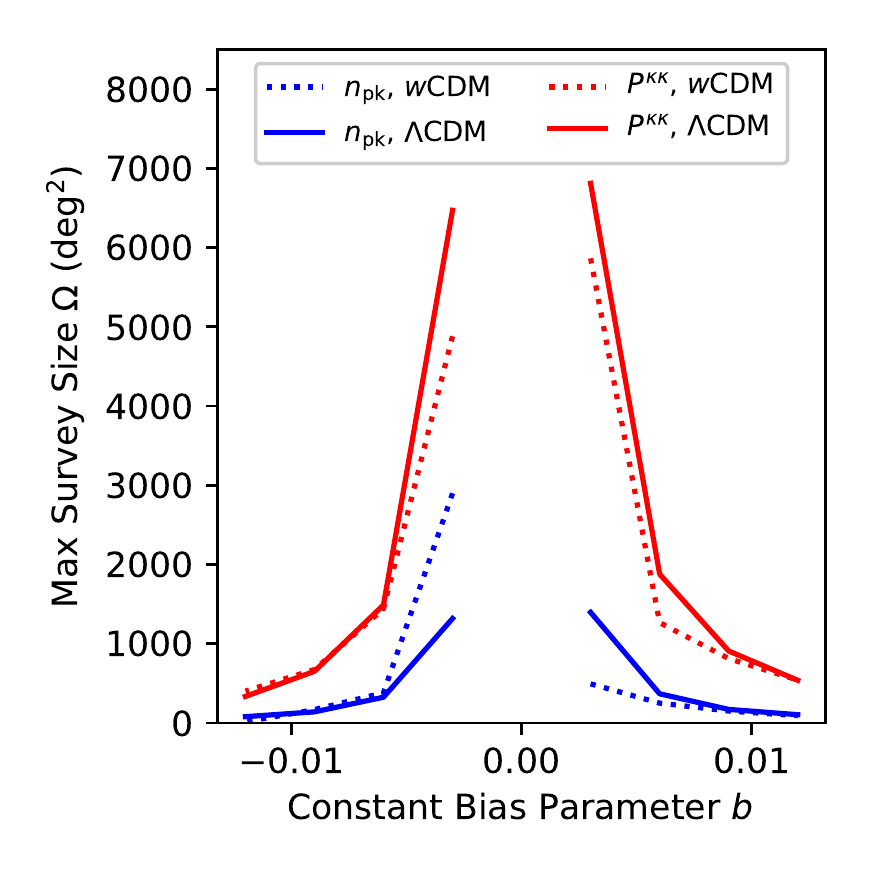}
\caption{\label{figure:survey_limits_bias} Maximum survey size for
  which photo-$z$ biases parameterised with constant parameter $b$
  produce tolerable error degradations. The lines on this plot roughly
  trace out the survey sizes for which the parameter bias is $50\%$
  larger than the distance to the $68\%$ confidence contour from the
  most-likely value along the direction of the bias. The four curves
  correspond to $\Lambda$CDM and $w$CDM parameter inferences from
  $n_{\rm pk}$ and $P^{\kappa\kappa}$, as labelled.}
\end{figure}

\section{Results}\label{section:results}

\subsection{Residual photo-$z$ Bias}
This work primarily focuses on the impact of residual
photo-$z$ bias on \estp. For discussions of our motivation and
how we model residual $b_{\rm ph}$, we refer to the reader back to
\S~\ref{section:m_zph_bias}. However, we remind the reader that 
each galaxy's spectroscopic redshift is given by its photo-$z$ 
subtracted by $b_{\rm ph}$. We divide our assessment of $b_{\rm ph}$ 
into two parts. First, we study a single source redshift and a simple constant $b_{\rm ph}$
to build our intuition. We then analyse a more realistic case involving 
tomography and an LSST-like \nph{z}.

\subsubsection{Single-$z$ dataset: photo-$z$ Biases}\label{section:simple_case}

We begin with the Single-$z$ dataset. 
The true distribution of galaxies and estimate are given
by $\ntrue{z}=25~{\rm arcmin}^{-2} \delta(z-1+b_{\rm ph})$ and $\nest{z} =
25~{\rm arcmin}^{-2}\delta(z-1)$.  Table~\ref{tab:simple_S8} lists the
values of $b_{\rm ph}$ we used. All values, other than
$b_{\rm ph}=0$, were selected such that each source plane lies just behind
a lens plane.

The left column of Figure~\ref{figure:simple_bias_hist_comp}
illustrates the impact of a bias $b_{\rm ph}\neq0$ on the peak counts in
the fiducial model. The top panel shows $n_{\rm pk}$ in the unbiased
($b_{\rm ph}=0$; blue dots), and a biased ($b_{\rm ph} = -0.086$; red
triangles) case. The bottom panel shows the fractional difference 
$[n_{\rm pk}(b_{\rm ph}=-0.086)-n_{\rm pk}(b_{\rm ph}=0)]/n_{\rm
  pk}(b_{\rm ph}=0)$.

The lower left panel of Figure~\ref{figure:simple_bias_hist_comp}
shows that outside of $0\la\nu\la2.5$, the fractional bias in the
number of peaks is proportional to $|\nu|$. For $2.5\la\nu\leq4.1$,
this finding is in qualitative agreement with that of
\citet{kacprzak16a} for shear peaks. Although our respective findings
differ for the number of peaks with $0\leq\nu\la2.5$, we attribute
this to the fact that \citet{kacprzak16a} examined relative
differences in the numbers of peaks after correcting for peaks
contributed by noise (which dominate at lower $\nu$), and that we each
use different $\ntrue{z}$ and filter scales.

We next contrast these changes in the peak counts with those
introduced by changes in cosmology. To this end, we define a quantity
often used to parameterise the degeneracy between $\Omega_m$ and $\sigma_8$,
\begin{equation}
  \label{eqn:S8}
  \Sigma_8 \equiv \sigma_8 \left(\frac{\Omega_m}{0.26}\right)^{\alpha}.
\end{equation}
We set $\alpha=0.59$ because it yields the most tightly constrained 
combination (for $b_{\rm ph}=0$) using the procedure outlined in \citet{petri15a}. 
The right panel in
Figure~\ref{figure:simple_bias_hist_comp} shows how $n_{\rm pk}$ is
modified in a cosmology with a higher $\Sigma_8$. This cosmology was
selected because it is among the \pset\ with the three closest
$\Sigma_8$ values to the best-fit value of the biased fiducial case
(shown in the left panel), and it has the most similar
$(\Omega_m,\sigma_8)$ to the fiducial values.  The side-by-side
comparison of the left vs. right panels illustrates that a negative
$b_{\rm ph}$ causes changes in $n_{\rm pk}$ that closely resemble the
changes arising from modifying $\Sigma_8$. This means that a negative
$b_{\rm ph}$ induces changes to $n_{\rm pk}$ similar to seeing a universe
with more evolved structure.
                                      
Figure~\ref{figure:shifted_contours} displays the $95.5\%$ confidence
contours of the inferred posteriors of \estp\ for a subset of $b_{\rm ph}$
values. The figure illustrates that a positive (negative) $b_{\rm ph}$ shifts
the posterior to the lower left (upper right) corner of the figure,
perpendicular to the direction of degeneracy, as expected from
Figure~\ref{figure:simple_bias_hist_comp}.

Because the contours largely retain their shape and only shift
perpendicular to the direction of degeneracy, we can quantify the
impact of $b_{\rm ph}$ on \estp\ by measuring the displacement of the
curve running along the direction of degeneracy, in the direction
perpendicular to the curves with $\Sigma_8=$constant.  For each value
of $b_{\rm ph}$, we compute the biased value of $\Sigma_8$ by plugging the
most-likely $\Omega_m$ and $\sigma_8$ into Equation~\ref{eqn:S8}.

For each $b_{\rm ph}$, Table~\ref{tab:simple_S8} lists the $\Sigma_8$ and
corresponding $\chi^2$ values inferred from the $n_{\rm pk}(\nu)$,
averaged over all 500 realisations of \kobs. Note that the negative
$\chi^2$ arise from our use of cubic interpolation. We also list the
corresponding averages, medians, and standard deviations
($\langle\Sigma_8^{r}\rangle$, median$\{\Sigma_8^{r}\}$,
std$\{\Sigma_8^{r}\}$) of the $\Sigma_8$ values fit in each individual
realisation of \kobs. 

Figure~\ref{figure:simple_S8_bias} displays the linear relationship
between $\delta\Sigma_8$, the bias in the posterior, and $b_{\rm ph}$. We
find the best-fit, when the $\delta\Sigma_8$ is defined as the
median$(\Sigma_8^{r})$ over all realisations. The best-fit line is
given by $\delta\Sigma_8 = (0.5211\pm0.0002) b_{\rm ph} + (0.01020\pm
3\times10^{-5})$. The non-zero $y$-intercept is artificial, and we
believe that it can be explained by the limitations of our
experiment. With more finely sampled cosmologies \pset, lens planes,
and $b_{\rm ph}$'s, the $y$-intercept should approach zero.

The above results can be interpreted intuitively as follows. If we
incorrectly assign lower redshifts to galaxies, we will be measuring
stronger lensing for this apparent redshift than we should be (since
the galaxies are farther away than we think). As a result, we will be
misled into thinking that our universe has a higher $\Sigma_8$ --
because in that case, the universe would have more evolved structure
today, causing a stronger lensing.

\subsubsection{Tomographic dataset: residual photo-$z$ biases}
\label{section:realistic_case}

Having built up basic intuition, we next assess the impact of $b_{\rm ph}$
on parameters inferred using tomography for a mock survey with an
LSST-like \nph{z} using the Tomographic dataset.  

Figure~\ref{figure:small_const_bias_contours} shows the $95.5\%$
confidence contours inferred for a $w$CDM cosmology from $n_{\rm pk}$
(left column) and from $P^{\kappa \kappa}$ (right column) from mock
observations of a single subfield with bias parameter $b$ values of
$\pm0.003$, $\pm0.006$, $\pm0.009$, and $\pm0.012$.
Figure~\ref{figure:small_const_bias_contours_N177} is identical except
that the contours are shown for a subset of the $b$ values for a
scaled survey including $N=177$ subfields ($\sim 2000~{\rm
  deg}^2$). The confidence contours for the scaled posteriors are
equivalent to high likelihood regions of the unscaled posteriors.
In both figures, the confidence contours in the top (bottom) row are 
drawn for posteriors marginalised over $w$ ($\sigma_8$), and are 
clipped due the finite extent of our non-zero uniform prior.

The shifts in the unscaled $95\%$ confidence contours in both the
($\Omega_m,\sigma_8$) and ($\Omega_m,w$) planes are smaller for
$P^{\kappa\kappa}$ than for $n_{\rm pk}$. While the changes in the
unscaled ($\Omega_m,\sigma_8$) contours inferred with
$P^{\kappa\kappa}$ are similar to the shifts discussed in
\S~\ref{section:simple_case} for the Single-$z$ dataset, the changes
in the contours inferred from $n_{\rm pk}$ are more complex and
include noticeable deformations in the overall shape. We believe that
this more complex behavior can be explained by the tomographic
information breaking the degeneracy between $b_{\rm ph}$ and
$\Sigma_8$. Such an explanation only works for redshift-dependent
$b_{\rm ph}$.

At the same time, the ($\Omega_m,\sigma_8$)
contours in Figure~\ref{figure:small_const_bias_contours_N177} from both
statistics shift {\em along} the direction of degeneracy. Positive
(negative) $b_{\rm ph}$ shifts these contours towards more negative
(positive) $\Omega_m$ and slightly decreases (increases) the 
enclosed area. The shifts in the ($\Omega_m,w$) contours due
to small, positive (negative) $b_{\rm ph}$ for each statistic in
Figures~\ref{figure:small_const_bias_contours} and 
\ref{figure:small_const_bias_contours_N177} are consistent with 
the entire unscaled posterior shifting
coherently toward negative (positive) $\Omega_m$; they also cause the
scaled up contour to contract (expand).

We also briefly discuss the effects of $b_{\rm ph}$ on the unscaled
$95.5\%$ confidence contour in the ($\Omega_m,\sigma_8$) plane when
$b=\pm0.025,$ $\pm0.05$, and $+0.1$. These contours have been omitted
from Figure~\ref{figure:small_const_bias_contours} but are shown
in Figure~\ref{figure:large_const_bias_contours} of
Appendix~\ref{app:large_bph}. For galaxies with $z_{\rm ph}=1$, these
$b_{\rm ph}$ are of similar magnitudes to those assessed in
\S~\ref{section:simple_case}. At these larger $|b_{\rm ph}|$, we find a
sharp departure in the behavior of the $n_{\rm pk}$ contours.  These
contours display larger shifts in the $\Omega_m$ direction, more
significant shape deformation, and less smooth overall dependence on
$b$. In contrast, the shifts in the $P^{\kappa\kappa}$ contours are
consistent with proportionally larger magnitude shifts along the
direction of degeneracy that the shifts illustrated in
Figure~\ref{figure:small_const_bias_contours}.

Figure~\ref{figure:bias/uncertainty} shows the dependence of the
relative bias figure of merit \relbias\ (defined in the previous
section) on $\Omega$, the number of subfields in a scaled up survey, and
the bias parameter $b$. Because of the weak sensitivity of either
statistic to $w$, we primarily focus on \relbias\ for inference in
$\Lambda$CDM cosmologies (top row). For completeness,
Figure~\ref{figure:bias/uncertainty} also includes the \relbias\ for
$w$CDM cosmologies (bottom row). Data for $n_{\rm pk}$
($P^{\kappa\kappa}$) are displayed in the left (right) column. The
vertical gray lines indicate $\Omega_{\rm max}$, the largest $\Omega$ for which
we can reliably forecast \relbias\ (see \S~\ref{section:scaled_size}),
and the horizontal dashed lines mark our 50\% benchmark for
significant posterior degradation (see \S~\ref{section:b/u}). Any
error in the estimation of the most-likely \estp\ due to the finite
sampling of the inference grid is small and only relevant to
\relbias\ for $|b|\sim0.003$ due to the small absolute size of the
bias in \estp.

There are two salient points revealed by
Figure~\ref{figure:bias/uncertainty}. First, for large $\Omega$, 
the \relbias\ of each summary statistic roughly follows a power-law
$\relbias \propto \Omega^{1/2}$. Such behavior is expected at large $\Omega$
because the posterior should resemble a Gaussian. Although such 
resemblance is not apparent in 
Figure~\ref{figure:small_const_bias_contours_N177} for a $w$CDM 
cosmology with $\Omega=2002~{\rm deg}^2$, it is evident for the contours of
$\Lambda$CDM with $\Omega=984~{\rm deg}^2$. The other takeaway is that the 
uncertainty in \estp\ is degraded by $\ga50\%$ at smaller survey sizes for 
$n_{\rm peak}$ than for $P^{\kappa\kappa}$.

Figure~\ref{figure:survey_limits_bias} summarises the survey area
$\Omega$ at which a given $b$ degrades the uncertainty in \estp\ by
$\sim50\%$. To construct the figure, we use linear interpolation and
extrapolation, in log-log space, using the points from
Figure~\ref{figure:bias/uncertainty} to determine the survey sizes
for which $\relbias=1.5$. With the exception of the inference of a
$w$CDM cosmology with $n_{\rm peak}$ when $b=-0.003$, we
exclusively use data satisfying $\Omega\leq \Omega_{\rm max}$. For that
particular case, we use the interpolated value between 
$\Omega=2002~{\rm deg}^2$ and $\Omega=2896~{\rm deg}^2$ because it yields 
a more conservative and realistic result.

The main conclusion from this last figure is that a residual bias 
$b_{\rm ph}$ with $|b|=0.003$ degrades
$\Lambda$CDM parameter estimates for surveys with $\Omega\ga
1300\ {\rm deg}^2$ ($\ga 6500\ {\rm deg}^2$) when using $n_{\rm peak}$
($P^{\kappa \kappa}$).  For $w$CDM, the corresponding limits are
$\Omega\ga 490\ {\rm deg}^2$ ($\ga 4900\ {\rm deg}^2$). It is worth
noting that this effectively corresponds to a pessimistic ``worst case
scenario'': It is highly unlikely that the residual $b_{\rm ph}$ would
match the upper limit of the LSST photo-$z$ requirements {\it and}
have the same sign at all $z$.

Finally, one may ask if the presence of bias can be discovered
from poor goodness-of-fit values. We find that the reduced $\chi^2$
values are much smaller than unity in all cases discussed in this
section, except for peak counts at the largest bias ($b=0.1$, yielding
reduced $\chi^2 = 0.66$), indicating that the biased best-fit models
remain good fits to the mock data.

\subsection{Realistic photo-$z$ Errors}
\label{section:realistic_photoz_r}

In this subsection, we shift our focus to assessing the biases in
\estp\ induced by directly approximating \ntrue{z} with some variation
of \nph{z}. We begin by investigating the impact of unmodelled
photo-$z$ errors, the errors from simply setting
$\nest{z}=\nph{z}$. We then examine the impact of errors in the shape
of $n^i_{\rm ph}(z)$ relative to the shape of $n^i_{\rm true}(z)$ in the
absence of centroid bias. The latter case assumes that in each
tomographic bin, a calibration process successfully.
adds a constant to $n^i_{\rm ph}(z)$ to make its centroid match that of \nttrue{z}.
We remind the reader that for these assessments, we adopt the photo-$z$ PDF 
from a simulated calibration set taken from \citet{rhodes17a}. This simulated dataset
accounts for LSST's photometric filters and uses a basic spectral
template to obtain photo-$z$'s, but does not meet the LSST science
requirements (see Appendix~\ref{app:photoz_perf}).

The un-optimised performance of the simulated photo-$z$ PDFs we adopted
and our heavy reliance on photo-$z$ point estimates make the results 
conservative. Setting \nest{z} equal to the distribution of point 
estimates, rather than computing it like a real surveys (e.g. stacking 
redshift posteriors), makes our \nest{z} worse estimates of \ntrue{z} 
than those in real surveys. Furthermore, performing ray-tracing with these point 
estimates, instead of randomly drawing redshifts based on \ntest{z}, contrasts 
with how real surveys constrain cosmology with two-point statistics. This 
choice makes our results more sensitive to errors: even if \nest{z} 
were identical to the \nttrue{z}, our results remain biased, since we 
do not assign the correct redshift to each galaxy. Consequently, our 
estimates in this subsection for the maximum survey sizes that avoid 
parameter degradation are conservative lower limits.

\begin{figure}
  \center
\includegraphics[width = 3.5 in]{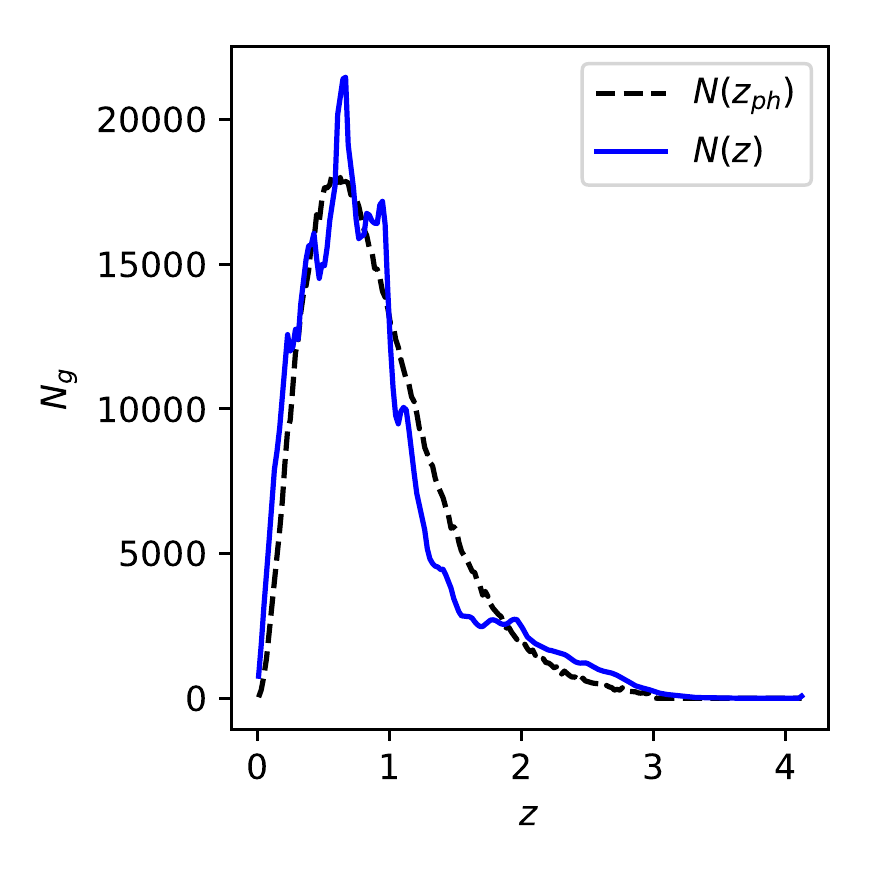}
\caption{\label{figure:realistic_n(z)} Illustrates the difference
  between the true galaxy redshift  distribution (blue solid line) and
  the galaxy photo-$z$ distribution (dashed black line) entering our 
  analysis. The true galaxy redshift distribution assumes that the 
  spectroscopic calibration sample follows the same distribution as 
  the surveyed galaxies.}
\end{figure}

\begin{figure*}
  \center
  \includegraphics[width = 6 in]{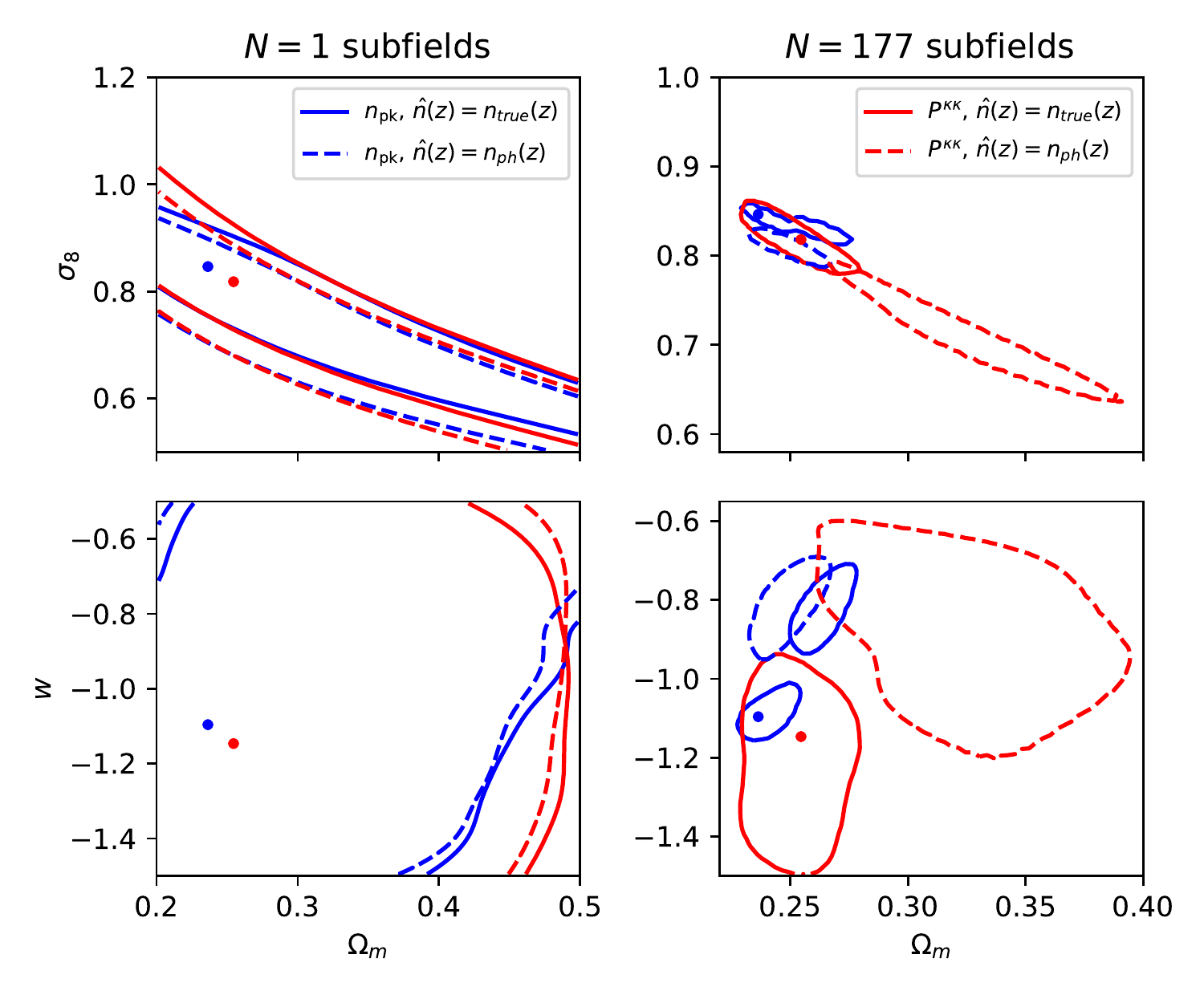}

  \caption{\label{figure:uncalibrated_photoz_bias}
    $95.5\%$ confidence contours from lensing peaks (blue) and from the
    convergence power spectrum (red). The left column shows the 
    unscaled contours while the right column shows the 
    contours scaled to $N=177$ subfields ($\sim 2000~{\rm deg}^2$).
    The solid lines correspond to the unbiased case, and the dashed lines
    to the case where $\ntrue{z}$ is approximated with $\nph{z}$ (i.e. 
    ignoring photo-$z$ errors), computed with the photo-$z$ PDF estimated 
    for LSST. The blue (red) points indicate the most likely cosmology inferred 
    from $n_{\rm pk}$ ($P^{\kappa\kappa}$) in the absence of photo-$z$ errors.}
\end{figure*}

\begin{figure}
  \center
\includegraphics[width = 3.5 in]{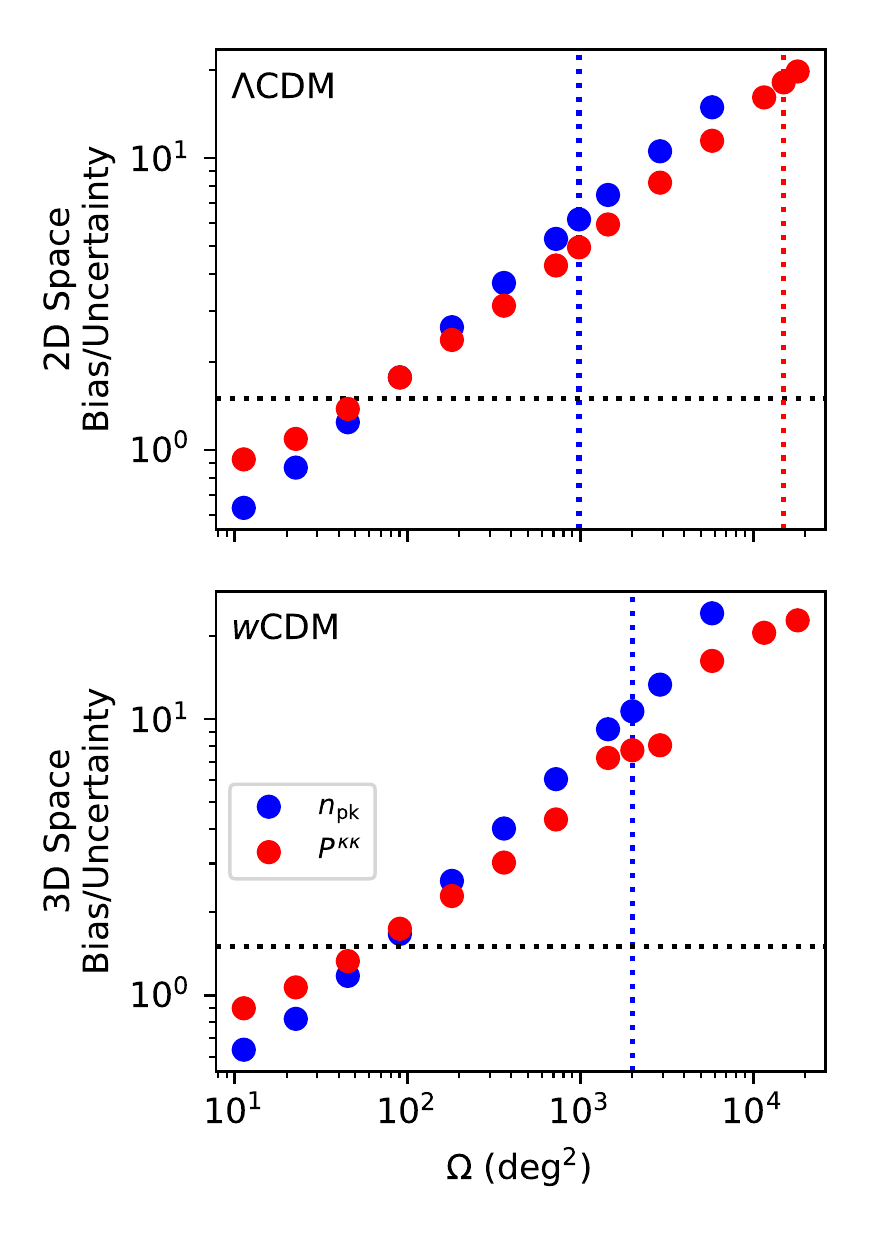}
\caption{\label{figure:uncalibrated_survey_limits}
  Biases in the inferred parameters relative to the distance to the
  $68\%$ confidence contour along the direction of bias (\relbias) as
  a function of scaled up survey size ($\Omega$). These biases are induced
  by approximating $\ntrue{z}$ with $\nph{z}$ for simulated LSST
  photo-$z$ performance. The top (bottom) panel shows results
  for a $\Lambda$CDM ($w$CDM) cosmology, and the blue (red) markers
  for $n_{\rm pk}$ ($P^{\kappa\kappa}$). The horizontal black dotted line
  corresponds to significant degradation (\relbias=1.5). The vertical blue (red) dotted lines
  indicate the maximum survey sizes above which numerical interpolation
  errors for $n_{\rm pk}$ ($P^{\kappa\kappa}$) invalidate our results.}
\end{figure}

\begin{table}
  \centering
  \caption{ Summary of inferred
    cosmological parameter biases arising from taking photo-$z$ point
    estimates at face value and assuming \nest{z}=\nph{z}. We list the
    reduced $\chi^2$ in each biased best-fit model, and the survey
    area $\Omega$ above which parameter estimates are degraded by
    $\geq 50\%$. The corresponding minimum average reduced $\chi^2$
    value of the unbiased $\Lambda$CDM constraints from $n_{\rm
      peaks}$ ($P^{\kappa\kappa}$) is 1.53 (1.14). These values
    were computed by averaging the sampled $\chi^2$ grids over $N_r$
    realisations.}
  \label{tab:calibration_neglection}
  \begin{tabular}{lcccc}
    \hline
    model & \multicolumn{2}{c}{$n_{\rm peaks}$} &
    \multicolumn{2}{c}{$P^{\kappa\kappa}(N_c=30)$} \\
    & reduced $\chi^2$ & $\Omega$ &
    reduced $\chi^2$ & $\Omega$ \\
  \hline
  $\Lambda$CDM & 1.15226 & $65.2\ {\rm deg}^2$ & 1.07622
  & $56.9\ {\rm deg}^2$ \\
  $w$CDM       & 1.16367 & $73.2\ {\rm deg}^2$ & 1.11603
  & $61.6\ {\rm deg}^2$ \\
  \hline
  \end{tabular}
\end{table}

\subsubsection{Unmodelled photo-$z$ Errors}
\label{section:calibration_error}

A simple way to frame the exercise in this subsection is that we are
examining the cosmological-parameter degradation arising from entirely ignoring
the photo-$z$ errors. Figure~\ref{figure:realistic_n(z)} illustrates
both \nph{z} (black dashed line) and the average true galaxy-redshift
distribution \ntrue{z} (blue solid line).

Figure~\ref{figure:uncalibrated_photoz_bias} 
qualitatively shows the shifts in the $95.5\%$ confidence contours for 
a single subfield (left column) and a $\Omega=2002~{\rm deg}^2$ survey 
(right column), for $w$CDM parameters. We remind the reader that the 
right column equivalently shows very high likelihood regions of the 
unscaled subfield. The figure shows sets of contours assuming 
$\nest{z} = \nph{z}$ (dashed lines) and $\nest{z} = \ntrue{z}$ (solid lines). The top
(bottom) rows shows the $\Omega_m-\sigma_8$ ($\Omega_m-w$) contours
marginalised over $w$ ($\sigma_8$). The
blue (red) lines illustrate the confidence contours inferred from
$n_{\rm pk}$ ($P^{\kappa\kappa}$).

Joint examination of the columns of Figure~\ref{figure:uncalibrated_photoz_bias}
indicate that the photo-$z$ errors induce a changes to the
$\Omega_m-\sigma_8$ posterior, inferred with $n_{\rm pk}$ and
marginalised over $w$, consistent with a coherent shift of the
posterior towards negative $\sigma_8$. The change to the $\Omega_m-\sigma_8$ 
posterior for $P^{\kappa\kappa}$ is different; the unscaled contour shifts 
towards negative $\Sigma_8$ while the high likelihood contour shifts and expands along 
$\Sigma_8$. We also find that the $\Omega_m-w$ contour inferred 
with $n_{\rm  pk}$ ($P^{\kappa\kappa}$), from a $\Omega=2002~{\rm deg}^2$ survey, 
shifts toward negative $\Omega_m$ (positive $\Omega_m$ and $w$). However, the 
changes to all $\Omega_m-w$ contours inferred from a single subfield, are ambiguous 
due to the large size of the contours 
compared to the prior. 

In Figure~\ref{figure:uncalibrated_survey_limits} we illustrate
\relbias\ as a function of $\Omega$. Results from $n_{\rm pk}$ 
($P^{\kappa\kappa}$) are shown using blue (red) markers and lines, 
and for $\Lambda$CDM ($w$CDM) cosmologies in the top (bottom) panel.
The solid vertical lines mark the largest numerically reliable $\Omega$ 
and the horizontal black dashed lines mark our benchmark for 
significant error degradation. 
As in \S~\ref{section:realistic_case}, we focus on
the $\Lambda$CDM case since the $w$ constraints are overall very weak. 
Figure~\ref{figure:uncalibrated_survey_limits} indicates that at large $\Omega$, 
\relbias\ has a rough power-law dependence on $\Omega$ for both summary statistics.
The logarithmic slope of this relation for $n_{\rm pk}$ is $\sim0.5$, while
for $P^{\kappa\kappa}$ it is slightly shallower.

Table~\ref{tab:calibration_neglection} summarises the survey sizes
$\Omega$ at which the uncertainties in \estp\ are degraded by
$\approx 50\%$. This degradation occurs for both summary statistics
and in both $\Lambda$CDM and $w$CDM cosmologies when
$\Omega\sim 60~{\rm  deg}^2$. As one might expect, the large, complex 
photo-$z$ errors studied in this subsection degrade the posteriors 
considerably more than small residual $b_{\rm ph}$. 

Table~\ref{tab:calibration_neglection} also includes the unscaled 
reduced $\chi^2$ values of the most likely biased values. Note that 
these values are not directly comparable to those mentioned in 
\S~\ref{section:realistic_case}, since they were not calculated by 
averaging the sampled $\chi^2$ grid computed for each realisation
When we construct the unbiased realisation-averaged $\Lambda$CDM 
posteriors in this way we find that the peak counts and power 
spectrum have minimum reduced $\chi^2$ values of 1.53 and 
1.14. We note that this difference in methodology does not
alter the location of the unbiased most-likely values and has no 
noticeable effect on the actual shape of the contours.

\begin{figure*}
  \center
\includegraphics[width = 5in]{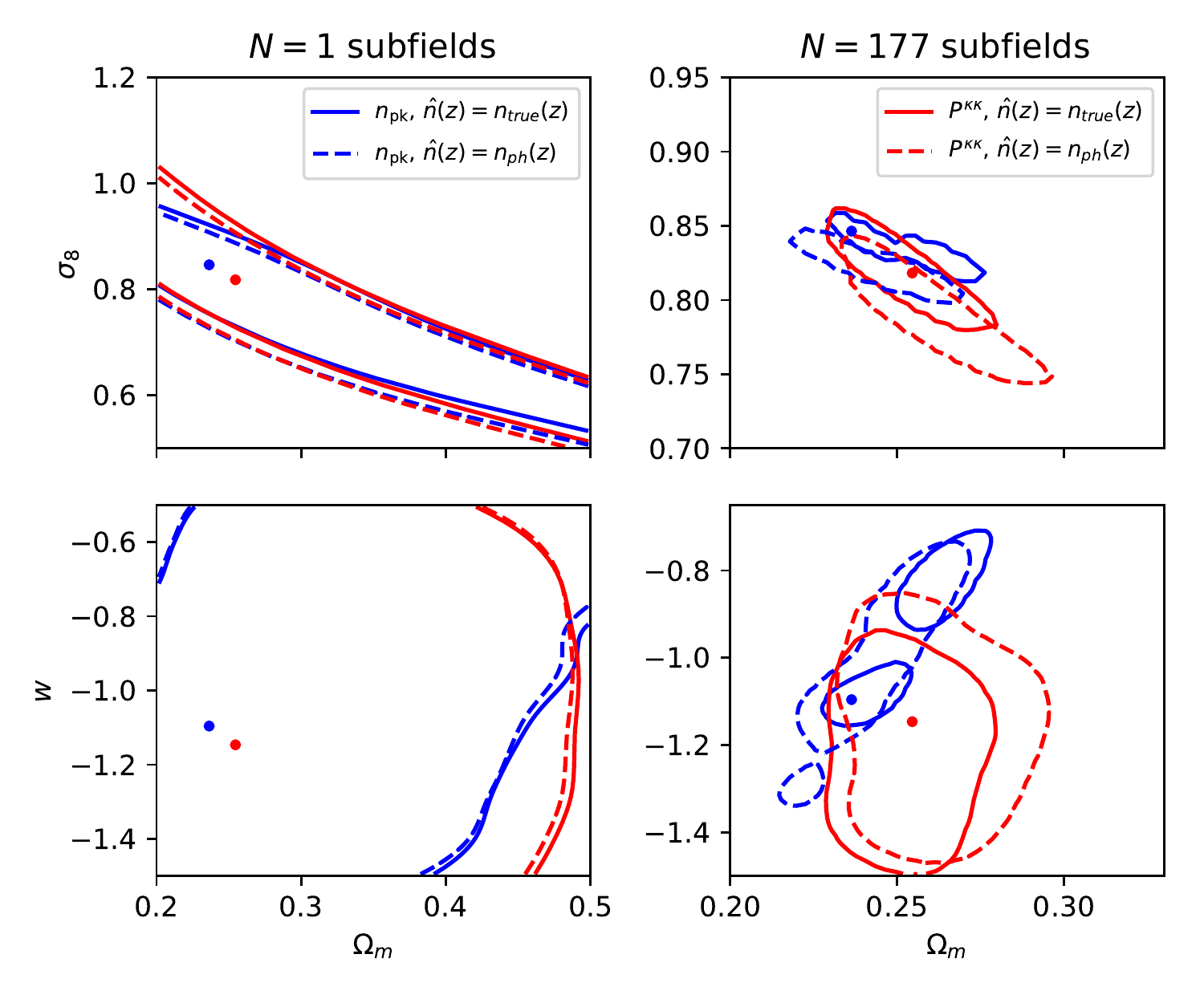}
\caption{\label{figure:no_cb_bias} Same as
  Fig.~\ref{figure:uncalibrated_photoz_bias} except that the
  the photometric redshift estimates (in each individual tomographic bin) have been
  modified by an additive constant, to remove the bias in their centroid in that bin.
Consequently, the biases 
illustrated in this figure arise mostly from differences between the shapes of 
the redshift distributions $n^i_{\rm true}(z)$ and $n^i_{\rm ph}(z)$ in each bin $i$. }
\vspace{\baselineskip}
\end{figure*}

\begin{figure}
  \center
\includegraphics[width = 3.5 in]{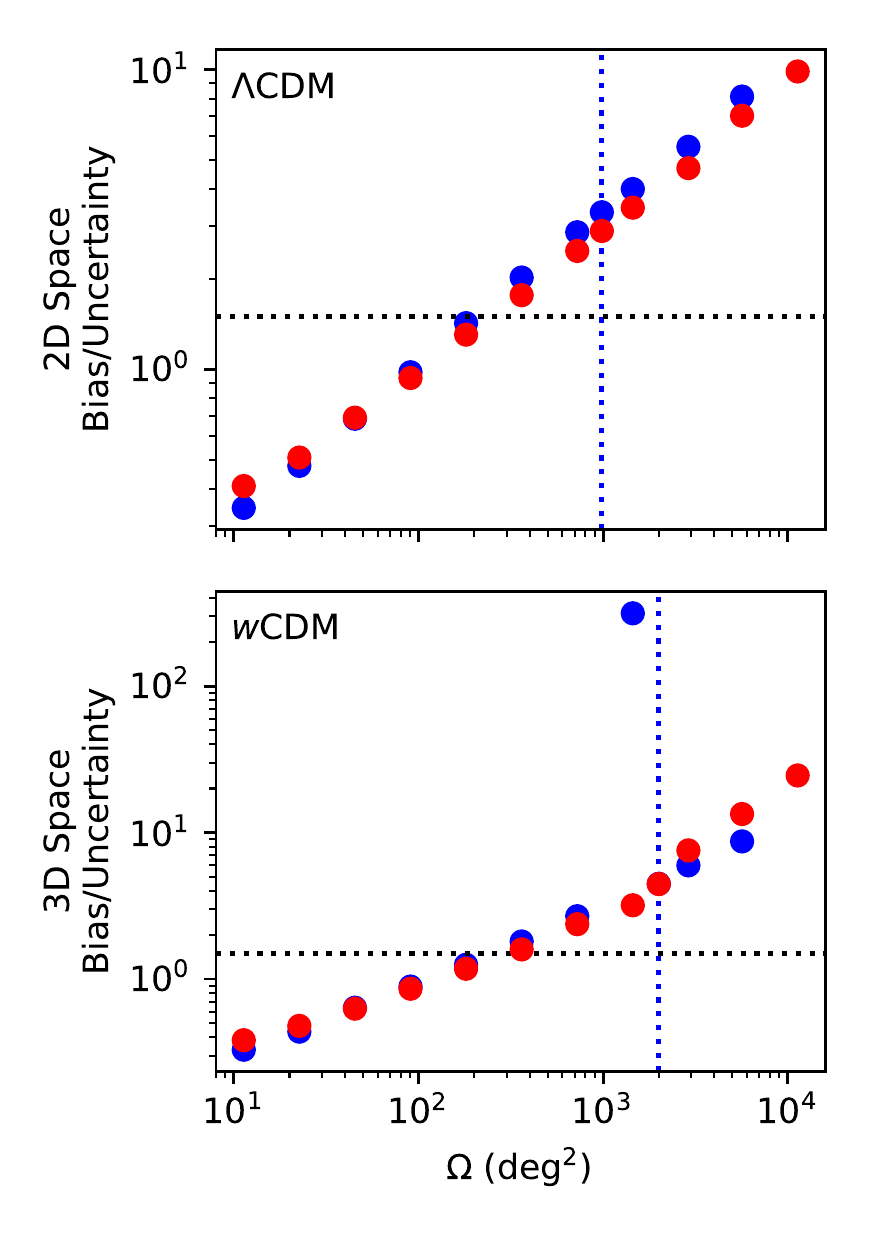}
\caption{\label{figure:no_cb_survey_limits} Same as 
Fig.~\ref{figure:uncalibrated_survey_limits}, except that the biases
arise from differences in the shapes of each the
redshift distributions $n^i_{\rm true}(z)$ and $n^i_{\rm ph}(z)$ in each
 tomographic bin $i$.}
\end{figure}

\begin{table}
  \centering  
  \caption{ Summary of cosmological parameter
    biases arising after the centroid bias of photo-$z$'s in each
    tomographic bin has been removed.  We list the average reduced
    $\chi^2$ values of the most likely values for a single subfield
    and the survey area $\Omega$ at which parameter estimates are
    degraded by $\approx 50\%$. For reference, the corresponding
    average reduced $\chi^2$ values of the unbiased best-fit models in
    a $\Lambda$CDM cosmology from $n_{\rm peaks}$ ($P^{\kappa\kappa}$)
    are 1.523 (1.14). }
  \label{tab:no_cb}
  \begin{tabular}{lcccc}
    \hline
     model & \multicolumn{2}{c}{$n_{\rm peaks}$} &
    \multicolumn{2}{c}{$P^{\kappa\kappa}(N_c=30)$} \\
    & reduced $\chi^2$ & $\Omega$ & reduced $\chi^2$ & $\Omega$ \\
    \hline
    $\Lambda$CDM & 1.21 & $200\ {\rm deg}^2$ & 1.09
    & $248\ {\rm deg}^2$ \\
    $w$CDM       & 1.22 & $255\ {\rm deg}^2$ & 1.13
    & $315\ {\rm deg}^2$ \\
    \hline
  \end{tabular}
\end{table}

\subsubsection{$n^i(z)$ Shape Errors}
\label{section:no_cb_results}

Finally, we examine the biases in \estp\ that arise from discrepancies in the 
shapes of $n^i_{\rm true}(z)$ and $n^i_{\rm ph}(z)$, assuming perfect removal of 
centroid bias. As explained above, this approach is equivalent to assuming 
that a calibration process was
able to obtain the correct centroid of the redshift-distribution in
each tomographic bin.  In practice,  we consider each tomographic redshift
bin, and shift the \nttrue{z} by a constant,
such that the centroid of the photometric and spectroscopic redshifts coincide.
Figures~\ref{figure:no_cb_bias} and \ref{figure:no_cb_survey_limits}
illustrate the same information as
Figures~\ref{figure:uncalibrated_photoz_bias} and
\ref{figure:uncalibrated_survey_limits}, but they now pertain to
errors in the shape of $n^i(z)$ (i.e. after the removal of centroid
bias). We note that in scaled-up surveys, there is a choice to remove
centroid biases either from individual subfields, or from the
aggregate of the $N=177$ subfields ($\Omega=2002~{\rm deg}^2$). In 
practice we find that this choice makes no discernible difference to 
the confidence contours.

The left column in Figure~\ref{figure:no_cb_bias} shows that the biases in
the $95\%$ confidence contours in a single subfield due to shape errors in $n^i(z)$ are 
roughly consistent with being smaller magnitude 
versions of the shifts that arise from directly approximating \ntrue{z}\ with \nph{z} (also for a 
single subfield). However, Figure~\ref{figure:no_cb_bias} shows that 
these trends do not continue for the biases in the highest likelihood regions
Instead, we find that the $\Omega_m-\sigma_8$ contours inferred from $n_{\rm pk}$ shift
perpendicular to their direction of degeneracy, toward negative $\Sigma_8$, 
but with some additional shift in the negative $\Omega_m$ direction. In contrast, the
$\Omega_m-\sigma_8$ contours from  $P^{\kappa\kappa}$ slightly stretch along their 
direction of degeneracy and shift toward negative $\sigma_8$ and positive
$\Omega_m$. Additionally, we find that the ($\Omega_m,w$) contours from
$n_{\rm pk}$ ($P^{\kappa\kappa}$) slightly shift in the negative
(positive) directions of both $\Omega_m$ and $w$. While the
($\Omega_m,w$) contours from $n_{\rm pk}$ expand in the direction of
their degeneracy, the $P^{\kappa\kappa}$ contours expand in all
directions.

Figure~\ref{figure:no_cb_survey_limits} indicates that $\approx50\%$
degradations occur for $\Lambda$CDM ($w$CDM) constraints inferred from
$n_{\rm pk}$ and $P^{\kappa\kappa}$ at survey sizes of 200 (255)
deg$^2$ and 248 (315) deg$^2$. The reduced $\chi^2$ values are
summarised in Table~\ref{tab:no_cb}. We remind the reader that the
corresponding reduced $\chi^2$ values for unbiased cases in
$\Lambda$CDM from $n_{\rm pk}$ and $P^{\kappa\kappa}$ are 1.53 and
1.14, respectively.

\section{Discussion}\label{section:discussion}

We begin this section by first discussing the biases of $\Sigma_8$
induced by $b_{\rm ph}$ for the Single-$z$ dataset. A simplistic toy
model, assuming that $\kappa$ peaks are entirely explained by lensing
from dark matter halos, illustrates why $b_{\rm ph}$ and $\delta\Sigma_8$ are
anti-correlated. If there is a negative (positive) $b_{\rm ph}$, an
observer will mistakenly believe that observed source galaxies are
closer to (further from) them than the galaxies are in reality. To produce a
given magnitude $\kappa$ value over a shorter (longer) distance, an
observer will infer that on average halos must be more (less)
massive. This is achieved in cosmologies with larger (smaller) values
of $\Sigma_8$. The same basic logic applies of course to other over-densities
(not just halos), and it also explains the
direction of shifts arising from residual $b_{\rm ph}$ in the
$\Omega_m-\sigma_8$ confidence contours in the Tomographic dataset, 
inferred from either $n_{\rm pk}$ or $P^{\kappa\kappa}$.
However, it does not explain the shifts in the most likely values.

A comparison of the shifts in the $\Omega_m-\sigma_8$ confidence
contours from $n_{\rm pk}$ between Figures~\ref{figure:shifted_contours}
and \ref{figure:small_const_bias_contours} illustrates that the
combined effects of tomography and $|b_{\rm ph}|\propto z$ induce
more complex changes in the posteriors of \estp. At the same
time, comparisons of the shifts in the contours of
Figure~\ref{figure:small_const_bias_contours} inferred by $n_{\rm pk}$
and $P^{\kappa\kappa}$ suggest that $P^{\kappa\kappa}$ is more resilient to
$b_{\rm ph}$ than $n_{\rm pk}$.

To assess the validity of \relbias\ as our proxy for error degradation,
we compare our findings, regarding the impact of $b_{\rm ph}$ on the
inference of a $w$CDM cosmology using $P^{\kappa\kappa}$, with those of
\citet{huterer06a}. Their analysis employs a Fisher Matrix formalism
to model degradations in the marginalised error of $\Omega_m$,
$\sigma_8$, and $w$ induced by residual centroid biases $\delta z$. 
\citet{huterer06a} define 10 tomographic bins, of equal
widths spanning $0<z_{\rm ph}<3.0$, and model the priors on $\delta z$ as
independent zero-mean Gaussians with identical variances $\Delta^2$.

The choices of tomographic bins, higher shape noise, and higher total
galaxy surface density made by \citet{huterer06a} cause their 2 (4)
tomographic bins at $z>2.4$ ($z<0.3$ and $1.5<z<2.4$) to have
values of $(\sigma_\gamma^2/n^i_g)$ that are factors of $\sim7-13$ 
($\sim1.3-4$) larger than the corresponding value averaged over all of 
our tomographic bins. Note that $n^i_g$ indicates the galaxy number 
density for tomographic bin $\mean{z}_b$. Because the covariance of 
$P^{\kappa\kappa}(\ell,\mean{z}_b,\mean{z}_b^\prime)$ and
$P^{\kappa\kappa}(\ell,\mean{z}_b,\mean{z}_b^{\prime\prime})$ is correlated 
with $(\sigma_\gamma^2/n^i_g)$, these choices lead \citet{huterer06a} to infer 
slightly weaker undegraded constraints and sensitivity of $P^{\kappa\kappa}$ to 
$\delta z$. Our use of each galaxy's photo-$z$ point estimate for ray-tracing
(rather than drawing from redshifts $\hat{n}^i(z)$) also augments this 
effect. At the same time, both their use of an LSST-like \ntrue{z} 
peaking at $z= 0.7$ (rather than 0.6) and their calculation of 
$P^{\kappa\kappa}$ over a $\sim60\%$ larger multipole range 
($50<\ell<3000$) has the opposite effect.

For 4900 and 6300 square degree surveys, \citet{huterer06a} finds
$50\%$ marginalised error degradations when the priors on $\delta z$
have $\Delta$ $\sim0.0019-0.0026$ and $\sim0.0017-0.0023$. For
simple comparison to our results, we assume that the error degradations
occur when $|\delta z| \sim \Delta$. At the same survey sizes, we
find $\approx 50\%$ degradation in the posterior when $b$ is $-0.003$
and $0.003$, respectively. In our analysis, $|b|=0.003$ corresponds to
$|\delta z|\sim0.0037-0.0089$, with larger magnitudes at larger
$\mean{z}_b$. At the $\mean{z}_b$ where the results of
\citet{huterer06a} are most relevant ($z\sim0.5-1$), we predict
$50\%$ degradations for effective centroid biases,
$|\delta z|\sim0.0046-0.0060$.  These values are reassuringly close
\citet{huterer06a}'s; i.e. factors of $\sim1.8-3.5$ larger than
their results.

We next turn our attention to
Figure~\ref{figure:survey_limits_bias}, which shows the maximum survey 
size at which error degradation from $b_{\rm ph}$ is tolerable, as a function 
of $b$. For the most pessimistic cases of $b_{\rm ph}$ that satisfy
the LSST photo-$z$ requirements ($|b|=0.003$), we find $\sim50\%$ error
degradations can occur in the inferred posteriors of a $\Lambda$CDM
cosmology using $n_{\rm pk}$ ($P^{\kappa\kappa}$) in surveys as small
as $\Omega\sim 1300\ {\rm deg}^2$ ($\sim 6500\ {\rm deg}^2$).
Figure~\ref{figure:survey_limits_bias} also reinforces the idea that
$P^{\kappa\kappa}$ is less sensitive than $n_{\rm pk}$ to $b_{\rm ph}$,
independent of our model cosmology type.

We next compare our results for uncalibrated photo-$z$ errors with those of 
\citetalias{petri16b}. Their analysis quantified the joint impact of 
both $b_{\rm ph}=0.003(1+z)$ and scatter in $z_{\rm ph}$ on \estp\ using 
the power spectrum and scaled peak counts with 5 tomographic bins. 
Their assessment assumed linear dependence of the (mean) summary statistic 
$\Bd(\Bp)$ on cosmology $\Bp$. Additionally, they held \nest{z} across 
tomographic bins equal to \ntrue{z} and only allowed photo-$z$ errors to reshuffle 
redshifts between tomographic bins (modifying \ntest{z}). Keeping these difference 
in mind, we compare the shifts in the ($\Omega_m,w$) contours of
Figure~\ref{figure:small_const_bias_contours_N177} arising from
$b=0.003$ with the distribution of biased most likely ($\Omega_m,w$)
values reported by \citetalias{petri16b} inferred from 20 semi-independent
mock observations of an LSST-sized survey area. 
Although our results basically agree on the directions of the biases, 
\citetalias{petri16b} finds that $P^{\kappa\kappa}$ is more sensitive to 
$b_{\rm ph}$ than $n_{\rm pk}$ and that unbiased cosmology is included within the
$68\%$ confidence contour of the distribution of the biased values inferred 
with $n_{\rm pk}$.

Though many of the differences in our analyses may contribute to this
discrepancy, one compelling explanation is our choice of tomographic 
bins. \citetalias{petri16b} attributes their finding of extra resilience of 
$n_{\rm pk}$ to photo-$z$ errors to the absence of spatial correlation in 
photo-$z$ errors coupled with the fact that peak locations are determined by 
the shapes of several neighboring source galaxies. Our choice of tomographic bins 
reduces our average surface density of galaxies per bin ($\sim0.4$ galaxies per pixel) 
to half that of \citetalias{petri16b}. Because we have fewer galaxies around each peak, 
our $n_{\rm pk}$ may be more sensitive to photo-$z$ errors. The sensitivity of 
$P^{\kappa\kappa}$ to $b_{\rm ph}$ probably has weaker dependence on our choice of bins 
because we compute cross-spectra for each combination of bins. If we were to repeat 
our analysis only using 5 tomographic bins, we expect $n_{\rm pk}$ to be have resilience to 
photo-$z$ errors better than or comparable to that of $P^{\kappa\kappa}$. We defer 
exploration of this to future work.

The shifts of the $95\%$ contours in Figure~\ref{figure:uncalibrated_photoz_bias} 
suggests that the lower sensitivity of $P^{\kappa\kappa}$ to $b_{\rm ph}$ 
does not extend to arbitrary photo-$z$ errors. Joint examination of
the biases induced by different classes of photo-$z$ errors, in
Figure~\ref{figure:small_const_bias_contours_N177}
and in the right columns of Figures~\ref{figure:uncalibrated_photoz_bias} 
and \ref{figure:no_cb_bias}, hints at the utility of employing $n_{\rm pk}$ 
alongside $P^{\kappa\kappa}$ for large  surveys ($\Omega \ga 2002~{\rm deg}^2$).
Figure~\ref{figure:small_const_bias_contours_N177} shows that the 
$\Omega_m-w$ contours inferred from the two statistics have slightly 
different dependences on $b_{\rm ph}$  while the other figures show that the 
biases on \estp\ are very different for each statistic for more
complex photo-$z$ errors. This suggests that the using
$n_{\rm pk}$ alongside $P^{\kappa\kappa}$ might allow for
self-calibration of photo-$z$ errors.

We next turn to the main results of
\S~\ref{section:calibration_error}: for surveys as small as
$\Omega\sim 60\ {\rm deg}^2$, with LSST-like \nph{z} and photo-$z$
error, the approximation of \ntrue{z} with \nph{z}
degrades the posterior of \estp\ by $\sim 50\%$.
This implies a very strong bias, as the existing surveys already cover
areas that exceed $60\ {\rm deg}^2$.  However, our result has no direct
implication for current surveys with comparable or better sensitivity,
such as DES Y1 \citep{troxel17a}, because these surveys already employ better estimates 
of \ntrue{z}. E.g. the estimation of \ntrue{z} by 
stacking the individual redshift posteriors for each galaxy (rather than using the 
photo-$z$ point estimates) and by trying to remove centroid biases. To remove the 
centroid bias, these surveys use calibration techniques to place relatively tight 
priors on these biases which they then use when simultaneously inferring 
the cosmological parameters, the sizes of the centroid biases, and other 
nuisance parameters. 

In \S~\ref{section:no_cb_results}, we assess the impact of errors in the shape of 
$n^i(z)$ assuming that centroid biases have been perfectly removed during the estimation of 
\nttrue{z} with \ntph{z}. This is analogous to how current surveys remove 
centroid biases from their \ntest{z}. We find $\approx 50\%$ error degradations to 
$\Lambda$CDM ($w$CDM) constraints using $n_{\rm pk}$ and $P^{\kappa\kappa}$ for surveys with 
$\Omega=200$ (250) ${\rm deg}^2$ and $\Omega=248$ (315) ${\rm deg}^2$. These results still do 
not apply directly to current surveys, as our unbiased case at 200 square degrees provides tighter 
constraints than DES Y1 \citep{troxel17a} and the first year of HSC \citep{hikage18a}, the current
surveys with the strongest constraints. The tighter undegraded constraints make our relative biases 
larger. Moreover, the fact that our \ntest{z} is based on the distribution of photo-$z$ point-estimates
means that the errors in our final \ntest{z} are larger than those of either survey. 

As explained in the beginning of \S~\ref{section:realistic_photoz_r},
the results of \S~\ref{section:calibration_error} are 
~\ref{section:no_cb_results} are each pessimistic. In reality, a survey with an 
LSST-like photo-$z$ distribution and galaxy distribution would not encounter 
considerable degradations until survey sizes somewhat larger than we quote.

Finally, we emphasise that this work made a number of simplifying assumptions. Throughout
our analysis involving the Tomographic dataset, we assumed that
\relbias\ is a good proxy for error degradation, rather than
propagating the uncertainties in the inferred $\hat{n}(z)$
distributions allowed by the photo-$z$ errors. We also assumed
that using the predicted LSST spectroscopic galaxy distribution as
the estimate for the photo-$z$ distribution would incur negligible errors.
In \S~\ref{section:calibration_error}, we did not attempt to infer the
MLE of \ntrue{z}. Additionally, we assumed that the $p(e_z|z_{\rm ph})$ of
the simulated survey and calibration set were equivalent and that
$p(e_z|z_{\rm ph})$ is well described by a histogram. Future work needs to 
improve on our
results by addressing these assumptions. Furthermore, it is worth 
assessing how the inclusion of Euclid photometry improves our survey 
limits; \citet{rhodes17a} showed using Euclid can improve the scatter 
and outlier fraction of the photo-$z$ PDF by a factor of 2 for $1.5<z<3$ 
and improved the scatter by $\sim30\%$ at other $z$.

\section{Conclusions}\label{section:conclusion}

Photo-$z$ errors are expected to be one of the leading systematic errors 
in future WL surveys. In this paper, we have assessed the impact of two 
classes of photo-$z$ errors on cosmological parameters inferred from 
tomographic peak counts and the tomographic power spectrum. Other sources 
of systematics not addressed in this work include intrinsic alignment, 
measurement errors (correlated PSF residuals; de-blending; 
shape measurements), and other theoretical errors (simulation accuracy, 
finite number of realisations, baryonic effects).

To assess the implications of these errors, we use ray-tracing
simulations with a simple approach of modelling photo-$z$ errors to 
produce mock shape catalogues in $\sim 10$ deg$^2$ subfields of an
LSST-like survey. We focus primarily on quantifying the
degradation of constraints for a $\Lambda$CDM cosmology (but also
consider a $w$CDM cosmology). We address the degradations from residual photo-$z$
biases and from a (mis)estimation of the true galaxy distribution \ntrue{z} 
with various \nph{z}, distributions of photo-$z$ point estimates, using 
LSST-like photometric measurements.

Our main findings can be enumerated as follows.
\begin{enumerate}[(i)]
\item The power spectrum is more resilient than peak counts to
  residual photo-$z$ biases. However, both summary statistics appear
  to be similarly sensitive to more complex alterations to the
  inferred galaxy redshift distribution.
\item Pessimistic cases of residual photo-$z$ biases that satisfy the
  LSST photo-$z$ requirements can significantly degrade constraints. They
  can degrade constraints from surveys with an LSST-like galaxy redshift
  distribution, utilising lensing peak counts (the convergence power
  spectrum), as small as $\Omega\sim1200\ {\rm deg}^2$
  ($\sim6300\ {\rm deg}^2$) by $\approx50\%$.
\item Generally, surveys with LSST-like galaxy redshift distributions 
  and LSST-like photo-$z$ only as small as $\Omega\sim60\ {\rm deg}^2$ 
  can directly approximate \ntrue{z}\ with \nph{z}\ without degrading 
  their constraints by $\ga50\%$.
 \item If such surveys successfully remove the biases in the centroid
   of the photo-$z$ distribution in each tomographic bin, these
   critical survey sizes, corresponding to $\approx50\%$ parameter
   degradation, increase to $\Omega\sim200\ {\rm deg}^2$ for lensing
   peaks, and to $\sim250\ {\rm deg}^2$ for the power spectrum.
 \item The last result implies that even without centroid biases, the
   width and large tails of the unoptimised predicted photo-$z$ 
   PDF can significantly bias parameters. This needs to be further 
   mitigated with more sophisticated approaches of estimating \ntrue{z}
   (such as stacking redshift posteriors).
\end{enumerate}

Future work is needed to more fully understand the impact of photo-$z$
errors on non-Gaussian summary statistics in an LSST-like survey. 
Such work should include the simultaneous inference of 
cosmological parameters and the underlying galaxy redshift 
distribution in order to propagate uncertainties in the true underlying 
galaxy distribution to the final constraints. It should also employ 
realistic photo-$z$ errors and include the impact of uncalibrated photo-$z$ errors.
Other related necessary work includes assessing the impact of photo-$z$
errors on the inference of more complex dark energy equations of state
with non-Gaussian statistics and assessing the efficacy of employing
multiple summary statistics for self-calibration of photo-$z$ errors.

\section*{Acknowledgements}

We thank Jason Rhodes and Shoubaneh Hemmati for sharing the
simulated LSST
photo-$z$ distributions
\citep{rhodes17a} in electronic form, and Jos\'{e}
Manuel Zorrilla Matilla for useful discussions.  To perform
simulations and reduce data we used the NSF XSEDE facility. This work was also supported by NASA ATP grant 80NSSC18K1093. For data
reduction and analysis, we used \numpy\ \citep{vanderWalt11a}, \scipy\ \citep{jones01a}, \pandas\ \citep{mckinney10a}, 
\astropy\ \citep{robitaille13a}, \matplotlib\ \citep{hunter07a}, 
and the Launcher utility \citep{wilson14}.




\bibliographystyle{mnras}
\bibliography{allref.bib}



\appendix

\section{Boundary Conditions} \label{app:bc}

In this appendix we briefly discuss the impact of the boundary
conditions during the Kaiser-Squire Transform and Gaussian smoothing.
Per the discrete convolution theorem \citep{press92a}, the
straight-forward application of the Kaiser-Squire Transform
effectively assumes a periodic boundary condition. \citet{lanusse16a}
points out that the equivalent real-space convolution kernel
falls off as the angular transverse distance squared and by zero
padding the shear map, we can mitigate the effects of the periodic
boundary condition. Upon comparing the peak count histograms
(see the end of this section) with and without zero-padding (the
Fourier transforms were performed on an array with 4 times as many
pixels in the zero-padded case), we conclude that the error
introduced by assuming a periodic boundary condition is negligible.

For the Gaussian smoothing, we compare the variability in the
relative differences between outlying $\kappa$ values arising
from different choices of boundary condition and the
global variability of $\kappa$ in a smoothed map. To measure
variability, we use the median absolution deviation ($MAD$). The
relative differences are computed for a periodic boundary condition
with respect to a boundary condition in which the values are
mirrored across the center of the last pixel (the first repeated
value is the second pixel from the edge). We find that $MAD$s are
comparable when considering a square annulus of $\kappa$ values 8
pixels away from the edge. 

\section{Simulated Realistic photo-$z$ Performance}\label{app:photoz_perf}

\begin{figure}
  \center
\includegraphics[width = 3.25 in]{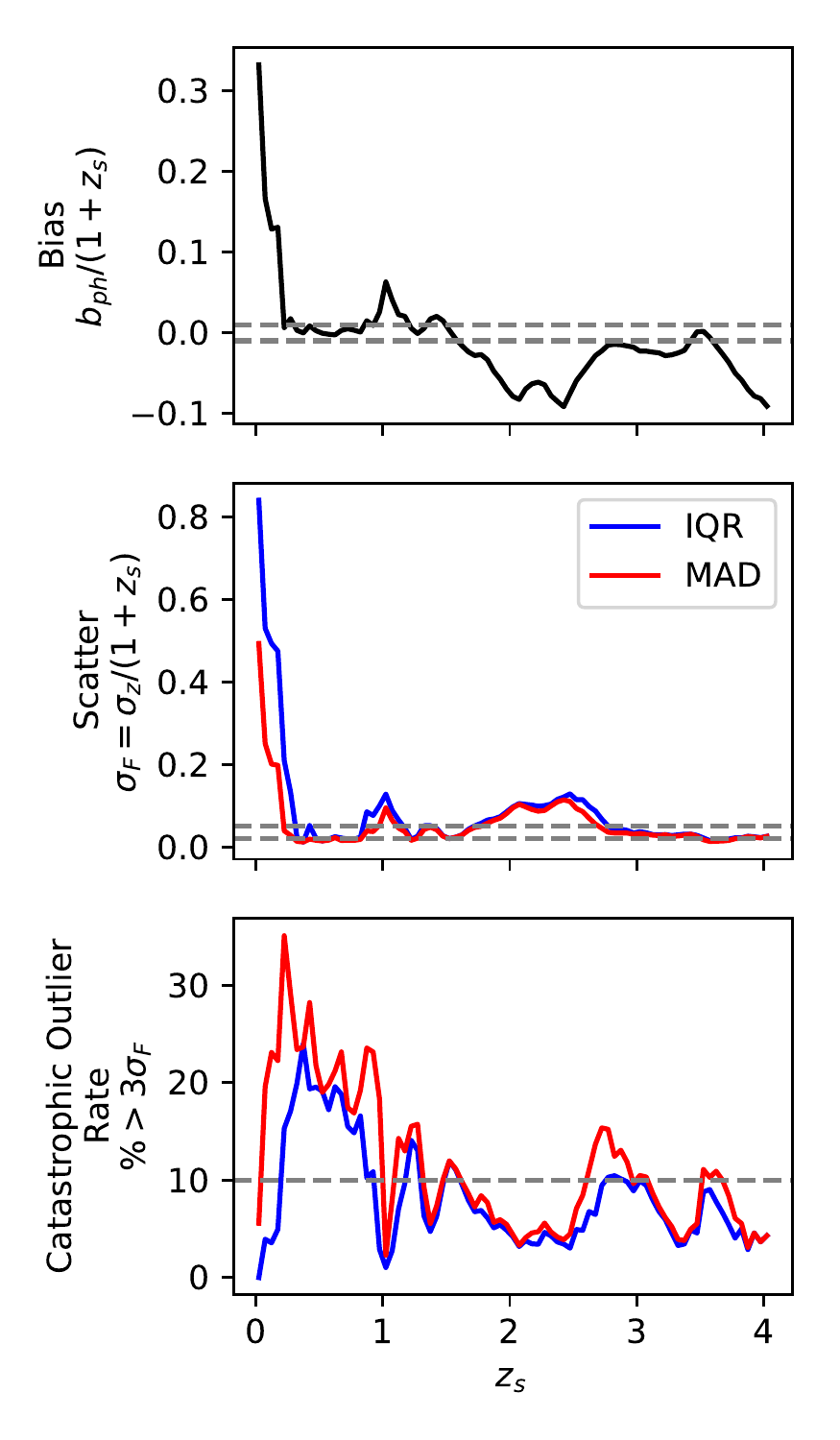}
\caption{\label{figure:photoz performance} Performance of realistic
  photo-$z$'s as a function of spectroscopic $z$ for bins with a width
  $\Delta z=0.05$. For each $z$ bin, we show the bias $b_{\rm ph}/(1+z)$,
  which is the median of $e_z=(z_{\rm ph}-z)/(1+z)$ (top), the scatter
  $\sigma_F$ about $b_{\rm ph}/(1+z)$ (middle), and the catastrophic outlier
  rate, defined here as the percentage of
  galaxies more than $3\sigma_F$ from bin. Specifically, the blue
  (red) lines indicate measurements when we define $\sigma_F$ as the
  interquartile range of $e_z$ divided by 1.38 (median absolute
  deviation of $e_z$ times 1.48). In each panel, the gray lines
  indicate the LSST photo-$z$ requirements \citep{lsst09a}. Due to the
  large magnitudes of $b_{\rm ph}/(1+z)$, the gray dashed lines shown
  indicate the joint weak lensing and galaxy clustering analysis
  requirement, $|b_{\rm ph}|<0.01(1+z)$, rather than the more strict
  requirement for weak lensing alone. The dashed lines in the middle
  panel indicate the strict requirement that $\sigma_z/(1+z)$ is less
  than 0.05 and illustrate the target value of 0.02.}
\end{figure}

In this appendix, we quantify the performance of the simulated LSST-like 
photo-$z$ measurements discussed in \S~\ref{section:m_realistic_error}. 
Figure~\ref{figure:photoz performance} shows the bias, scatter, and 
catastrophic error rate as a function of the true spectroscopic rate. The 
figure clearly shows that for each of these metrics, the realistic photo-$z$ 
measurements do not meet the LSST photo-$z$ requirements outlined in 
\citet{lsst09a}. 

Additionally, we compare our simulated photo-$z$ performance to that of 
the DES Y1 analysis, which currently provides one of the tightest sets of 
weak lensing \citep{troxel17a}. To do this, we measure metrics of the 
distributions of $R$, the difference between the photo-$z$ point estimate 
and its true spectroscopic redshift, for each tomographic bin. 
Specifically, we measure $\sigma_{68}(R)$, which is the half-width of 
the span of $68\%$ of values spanning the median of $R$, and the 
outlier fraction, which measures the fraction of $R$ values that deviate 
by more than $2\times \sigma_{68}(R)$ from the median of $R$. In 
Table~\ref{tab:photoz_details} we list each of these metrics computed for 
a single subfield averaged over the 1000 realisations of \scobs\ used 
to asses the impact of unmodelled photo-$z$ error.

The measurements listed in Table~\ref{tab:photoz_details} are directly 
comparable to those listed in Table 3 of \citet{hoyle18a} for the BPZ 
${\sc metacalibration}$, the same photo-$z$ distribution used to infer 
DES Y1 cosmology constraints \citep{troxel17a}. Our 8 lowest redshift 
bins each have a smaller $\sigma_{68}(R)$ than any of the 
bins employed for DES Y1, which are each about $\sim$0.14. However, all of our 
bins have outlier percentages that are larger by a factor of $\sim$2$-$6.
We note that our definition of $R$ slightly 
deviates from that of \citet{hoyle18a} due to our knowledge of the 
true spectroscopic redshift of each galaxy in our mock survey. We also 
note that the DES Y1 analysis only used these point estimates for 
identifying each galaxy's tomographic bin. The $\hat{n}^i(z)$ they use 
for cosmological inference was actually constructed by stacking the 
full the redshift posteriors produced for each galaxy by their 
photo-$z$ method \citep{hoyle18a}. Consequently, their estimates of 
$n^i_{\rm true}(z)$ are likely better than ours. 

\begin{table}
  \centering
  \caption{ Performance metrics for the 
  simulated realistic photo-$z$ 
  bin measurements from \citep{rhodes17a}, averaged over 1000 realisations. 
  Each metric is computed for the distribution of residuals $R=z_{\rm ph}-z$ in 
  a given tomographic bin. We include $\sigma_{68}(R)$, which is half of the 
  span of $68\%$ of the $R$ values centered on the median of $R$, and 
  Outlier Percentage, which is the percentage of values more than 
  $2\sigma_{68}(R)$ from the median residual.}
  \label{tab:photoz_details}
  \begin{tabular}{ccc}
    \hline
    $z_{\rm ph}$ range & $\sigma_{68}(R)$ & Outlier Percentage \\
    \hline
    $[0,0.332)$     & $0.02171\pm 0.00017$ & $20.11\pm 0.09$ \\
    $[0.332,0.464)$ & $0.0348\pm 0.0003$   & $22.36\pm 0.08$ \\
    $[0.464,0.577)$ & $0.0438\pm 0.0004$   & $18.90\pm 0.09$ \\
    $[0.577,0.689)$ & $0.0359\pm 0.0004$   & $19.33\pm 0.08$ \\
    $[0.689,0.806)$ & $0.0409\pm 0.0003$   & $16.63\pm 0.08$ \\
    $[0.806,0.936)$ & $0.0616\pm 0.0005$   & $17.24\pm 0.08$ \\
    $[0.936,1.089)$ & $0.1031\pm 0.0007$   & $18.65\pm 0.08$ \\
    $[1.089,1.287)$ & $0.1395\pm 0.0007$   & $17.67\pm 0.09$ \\
    $[1.287,1.596)$ & $0.2772\pm 0.0014$   & $12.82\pm 0.11$ \\
    $[1.596,3]$     & $0.310\pm 0.003$    & $14.91\pm 0.13$ \\
    \hline
  \end{tabular}
\end{table}

\section{Interpolation Comparison} \label{app:interp_comp}

\begin{figure*}
  \center
\includegraphics[width = 6.in]{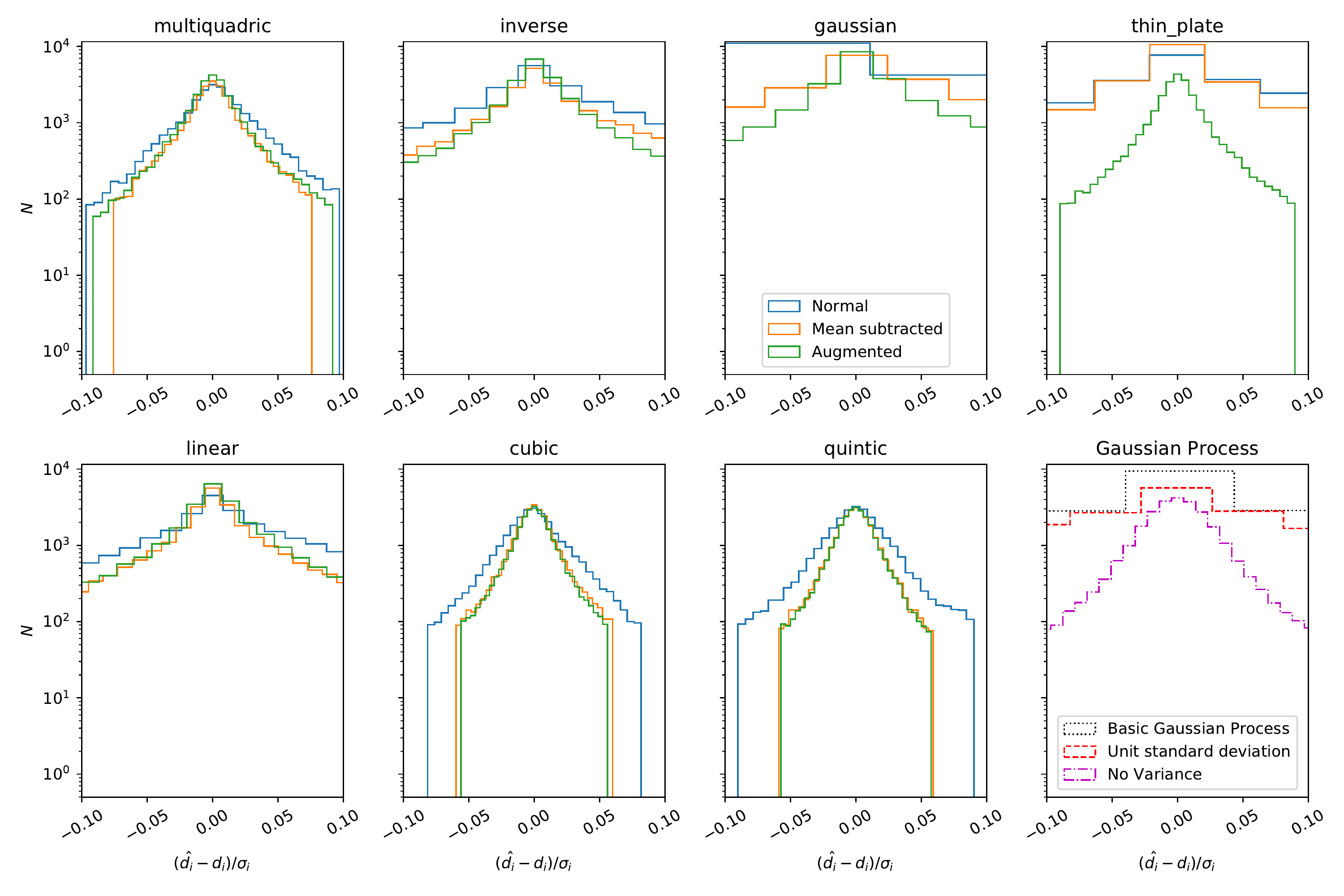}
\caption{\label{figure:sig_dev_95th_finite_percentile} Accuracy of
  different methods for interpolating summary statistics between
  cosmologies. Each panel shows the distribution of the $95\%$ lowest
  magnitude differences between predicted and true values relative to
  the standard deviation of each component of a summary statistic,
  with $\sigma\neq0$, for all 100 sampled cosmologies of the Shape
  Catalog Dataset. The summary statistic used to generate this figure
  is $n_{\rm pk}$ with 30 $\nu$ bins per tomographic peak. Every panel
  other than the lower right shows the performance of RBF
  interpolation (blue) and augmented RBF interpolation (green) using
  the unaltered components of $n_{\rm pk}$ with a different a basis
  function. In those panels we also show the performance of RBF
  interpolation for the components of $n_{\rm pk}$ subtracted by their
  cosmology averaged values (orange). The lower right panel shows the
  performance of Gaussian Process interpolation. Specifically, the
  panel shows the histograms for unaltered components using
  interpolation that accounts (does not account) for the variance in
  black (red) and for components rescaled such that the collection of
  a given component for all sampled cosmologies has unit standard
  deviation.}
\end{figure*}

Here we describe the accuracies of different schemes to interpolate a
summary statistic ${\bf d}$ as a function of cosmology. The emulator
interpolates $d_k$, the $k^{\rm th}$ component of a summary statistic,
at an arbitrary cosmology using the expectation values
$\{d_{k,1},d_{k,2},...,d_{k,P}\}$ in $P$ nearby sampled cosmologies
$\{{\bf p}_1,{\bf p}_2,...{\bf p}_P\}$.  The interpolation of each
component is performed independently.

By default, the emulator framework provided by \lenstools\ use
Radial Basis Function (RBF) interpolation. For some basis
function basis function $\phi(r)$, where
$r=\sqrt{{\bf p}^2-{\bf p^\prime}^2}$ is the distance between
two cosmologies in parameter space, the emulator predicts
the $k$th component as
\begin{equation}
  \label{eqn:rbf_interp}
  \hat{d}_k({\bf p}) = \sum_i^P\lambda_i\phi(\sqrt{{\bf p}^2-{\bf p}_i^2}).
\end{equation}
The values of $\{\lambda_1, \lambda_2, ..., \lambda_P\}$ are the solutions
to the system of equations obtained by plugging $\{{\bf
  p}_1,{\bf p}_2,...{\bf p}_P\}$ and $\{d_{k,1},d_{k,2},...,d_{k,P}\}$
into Equation~\ref{eqn:rbf_interp}.

\citet{liu15a} found that using a ``multi-quadric'' basis
function, $\phi(r) = \sqrt{(r/\epsilon)^2+1}$ ($\epsilon$ is a
constant typically set to the average distance between every pair of
sampled cosmologies) gave the most accurate results. However, our use of 
PCA to reduce
dimensionality complicates our choice of interpolation method. The
whitening step of PCA involves the subtraction of a constant value
$C$, the average value of $\{d_{k,1},d_{k,2},...,d_{k,P}\}$, from
the $d_k$ of every sampled cosmology. We define
$d_{k,\ell}^\prime =(d_{k,\ell}-C)$. Assuming that $d_{k}$ is not constant
for every sampled cosmology and that $C$ is non-zero,
$(d_{k,\ell}-C)/d_{k,\ell}$ is not constant for all $\ell$.
Consequently, RBF interpolation can predict different values for
$d_{k,\ell}^\prime+C$ and $d_{k,\ell}$ at a cosmology that was not sampled.
To help address this issue, we introduce another interpolation
technique called augmented RBF interpolation \citep{wright03a}.  This
method involves simultaneously performing RBF and linear
interpolations. The addition of a constant by the linear interpolation
makes this technique's predictions invariant to constant offsets of
$d_k$.

At this point we turn our attention to comparing the relative accuracy
of different interpolation methods. To do this, we select a cosmology
${\bf p}_j$ of the Tomographic dataset from the \pset\ and predict
$\hat{d}_k$ in this cosmology
with an emulator constructed from \pset\ but excluding
${\bf p}_j$. For this exercise, we choose to use the tomographic peak
count histogram, constructed with 30 equally sized $\nu$ bins in each
tomographic bin spanning the ranges given in Table~\ref{tab:tomo_details}.
We then compute $(\hat{d}_k-d_k)/\sigma_k$ where $\sigma_k^2$ is the
variance of $d_k$, computed from all $N_{r}=16000$ realisations of
$\kmap{_r}{,\bar{z}_b; \Bp_j}$.  We repeat this process for all $N_{b}=300$
components of \Bd\ and each ${\bf p}_j$ in \pset.

Figure~\ref{figure:sig_dev_95th_finite_percentile} shows histograms of
the $95\%$ lowest magnitude values of all $(\hat{d}_k-d_k)/\sigma_k$
with non-zero $\sigma_k$ for different interpolation methods. Each
panel, other than the bottom right, shows histograms for a given
radial basis function, offered by \scipy\ \citep{jones01a}. In these
panels, the green (blue/orange) line indicates the histogram using
augmented RBF interpolation (normal RBF interpolation on $d_k$ / on
$d_k-C$). The bottom right corner shows the performance for a naive
application of the Gaussian Process interpolation implemented by
\scikitlearn\ \citep{pedregosa11a}, with and without including the
$\sigma_k^2$ in the interpolation. It also shows the performance of
Gaussian Process interpolation applied such that the collection of
values of a given component of the summary statistic for all sampled
cosmologies has a standard deviation of 1.

Based on these results, we chose to use augmented RBF interpolation
with cubic basis functions ($r^3$), which gives the second-best
results in Figure~\ref{figure:sig_dev_95th_finite_percentile}.  While
quintic basis functions ($r^5$) yields marginally more accurate
results (i.e. a somewhat narrower distribution than the cubic case),
we chose cubic basis functions because they are simpler and the
difference is barely perceptible. Future work on this topic should
assess the impact of the relative scaling of different cosmological
parameters in distance calculations and assess the performance of more
carefully tuned Gaussian Process interpolation.

\section{PCA Application}\label{app:PCA}
\begin{figure*}
  \center
\includegraphics[width = 5. in]{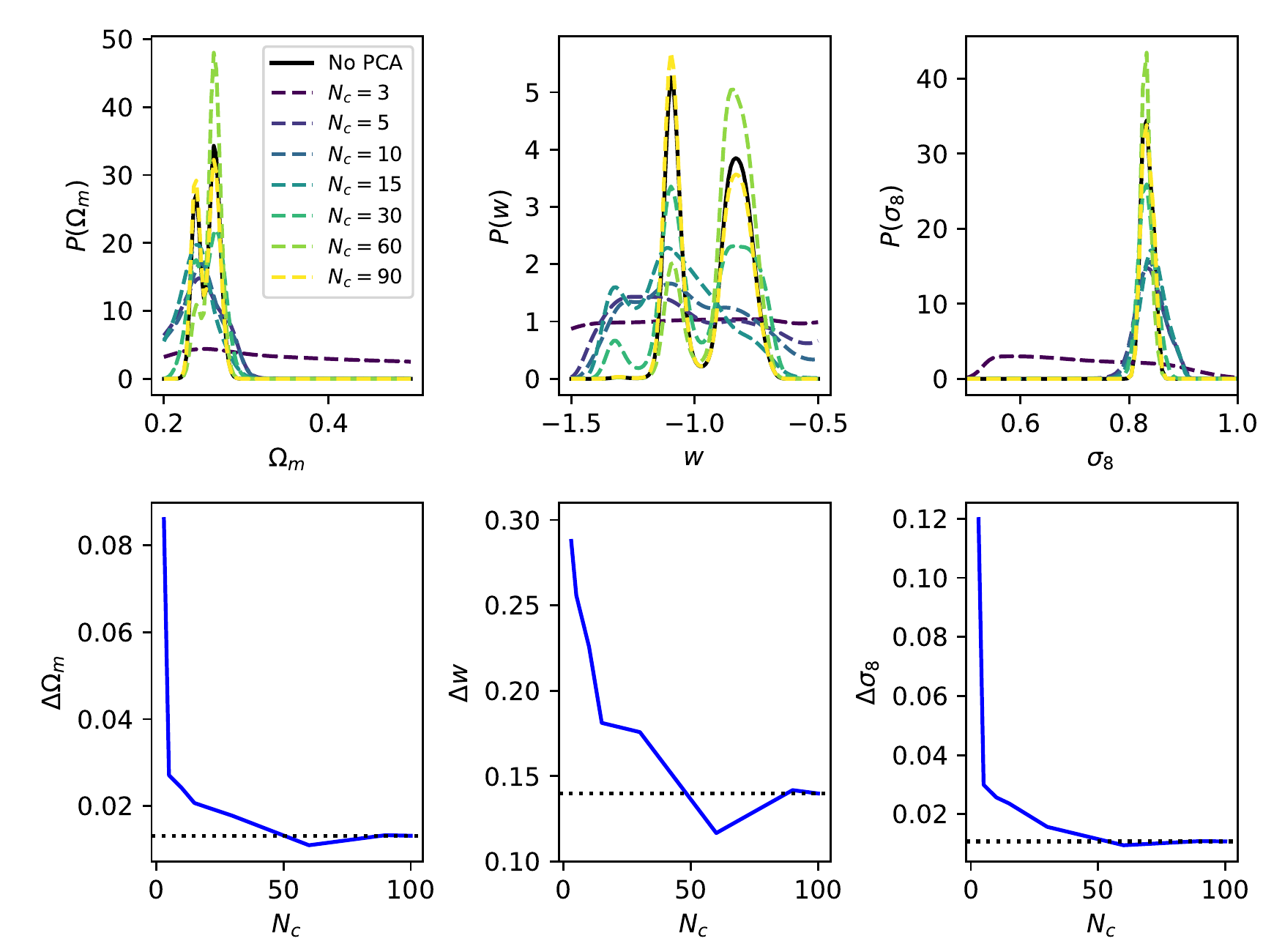}
\caption{\label{figure:pc_pca} The top row illustrates the
  marginalised posteriors for $\Omega_m$ (left), $w$ (center),
  $\sigma_8$ (right) achieved with different numbers of principal
  components ($N_c$) of the peak count histogram. Panels in the bottom
  row show the evolution of the standard deviations of the relevant
  marginalised posteriors as a function of $N_c$. The dotted black
  line shows the standard deviation of the full peak count histogram
  without any PCA. The data illustrated in this plot has been scaled
  up to a survey with $N=123$ fields.}
\end{figure*}

\begin{figure*}
  \center
\includegraphics[width = 5. in]{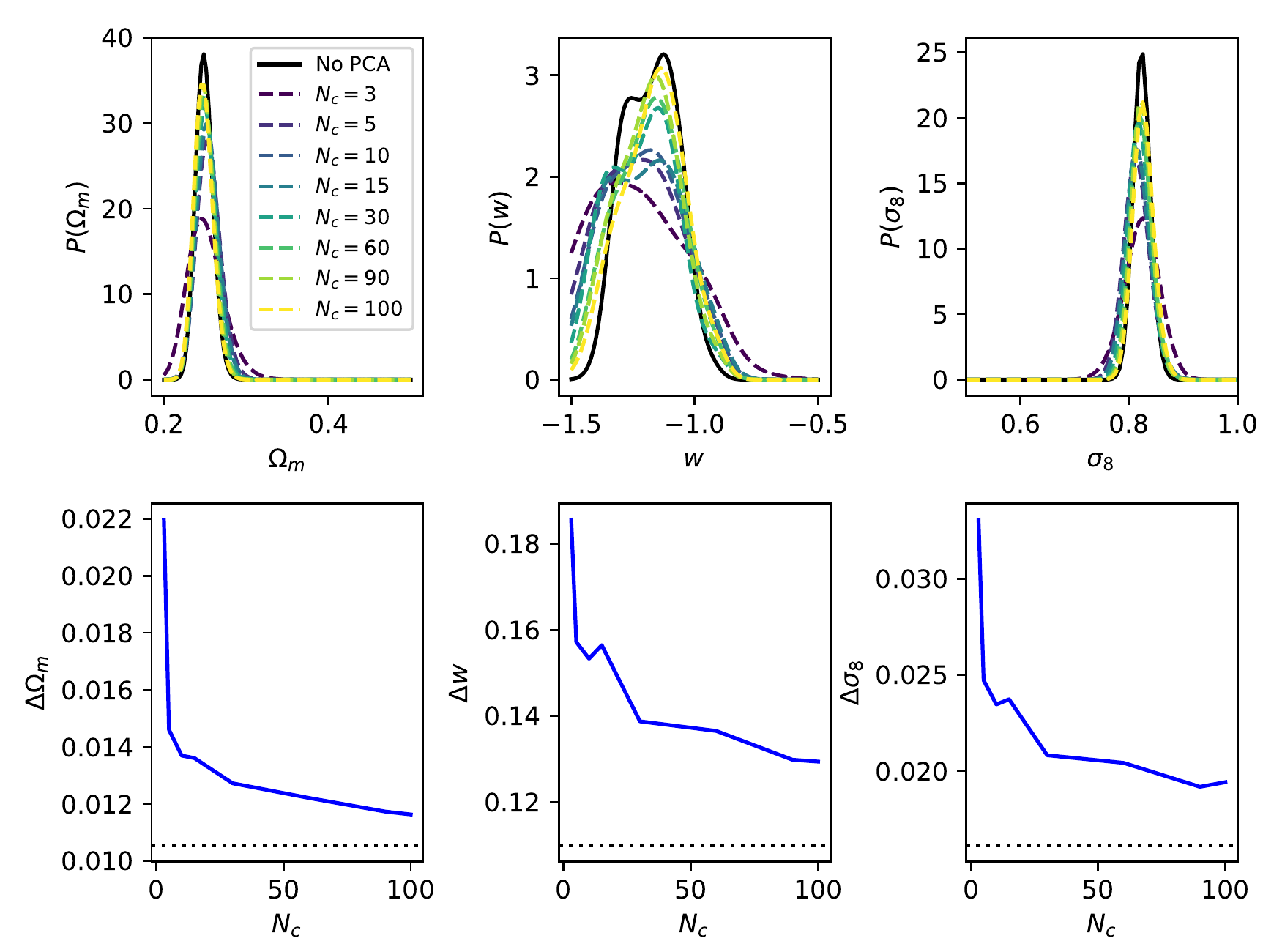}
\caption{\label{figure:ps_pca} Same as Figure~\ref{figure:pc_pca}
  except for the power spectrum. }
\end{figure*}

In this appendix we describe the process to identify the number of
principal components, $N_c$ used to represent each summary statistic.  
To do this, we examined the
change in the marginalised standard deviations of each of the
parameters in $w$CDM for a survey scaled up to $N=123$ subfields
(the maximum size allowed for the full power spectrum). The top row of
Figure~\ref{figure:pc_pca} (Figure~\ref{figure:ps_pca}) illustrates
the marginalised posteriors for different numbers of principal
components of the tomographic peak count histograms (tomographic power
spectrum) while the bottom row illustrates the change in the
marginalised standard deviation as functions of $N_c$. We found
negligible difference in employing the standard deviation from using
the size of the $68\%$ confidence interval.

Because of the dip in the marginalised standard deviations for the
peak counts when using $N_c=60$ below the marginalised standard
deviations using the entire feature set we elect to use the full peak
count histogram histogram in our analysis to avoid
misleading results. (This is equivalent to using $N_c=100$). Since the
rate at which the marginalised standard deviation decreases, largely
levels off at $N_c=30$, we elect to use $N_c=30$ principal components
for the power spectrum. 

\section{Larger Residual Bias}~\label{app:large_bph}

\begin{figure*}
  \center
\includegraphics[width = 4.5in]{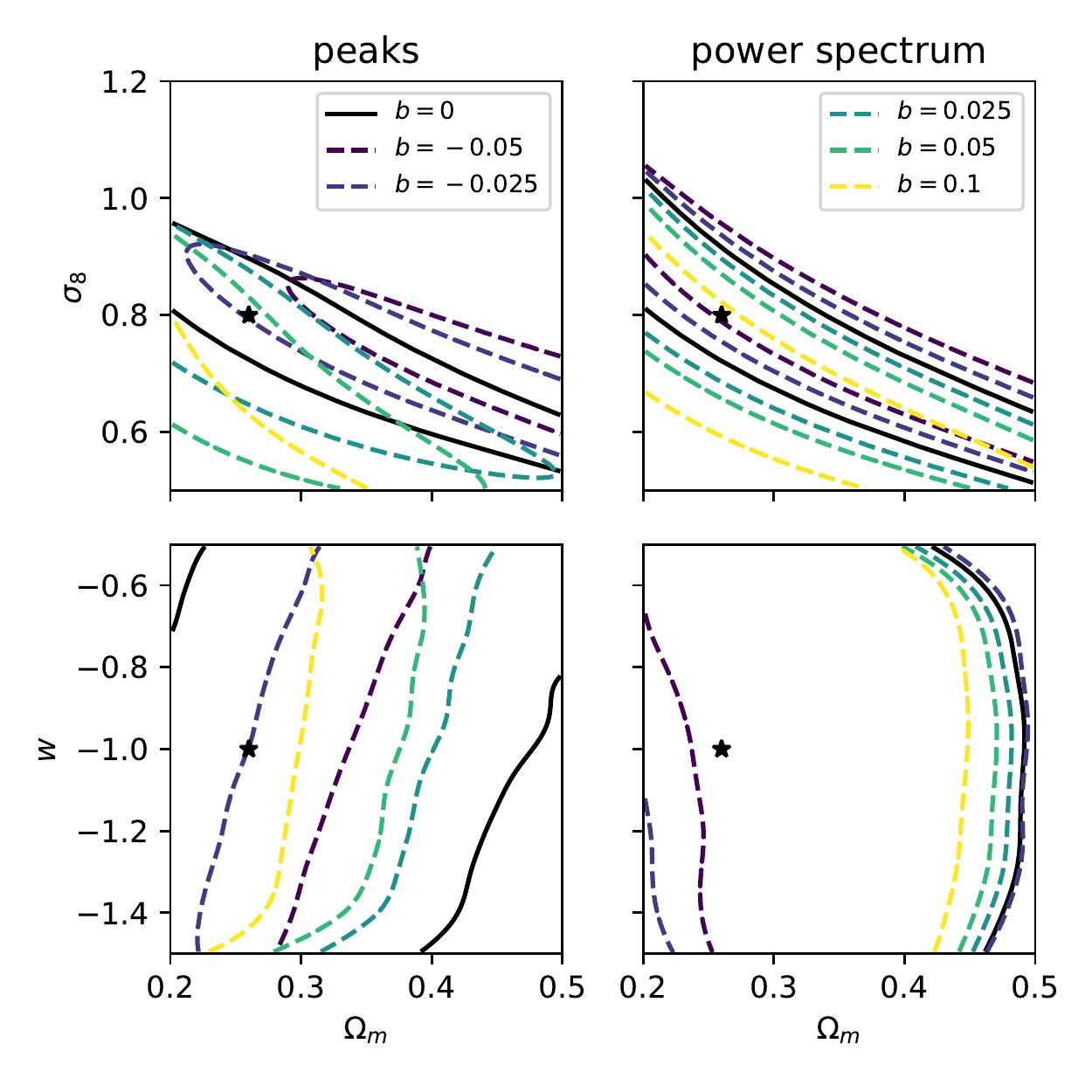}
\caption{\label{figure:large_const_bias_contours} Same as
  Figure~\ref{figure:small_const_bias_contours}, except that the 
  residual biases have much larger magnitudes.}
\end{figure*}

This appendix includes Figure~\ref{figure:large_const_bias_contours}
which illustrates the $95\%$ confidence contours which arise from
larger residual biases than considered in the main text. These
residual biases are also considerably larger than those allowed by the
LSST science requirements and are shown here for completeness.

\bsp	
\label{lastpage}
\end{document}